\documentclass{aa}
\usepackage{amssymb,amsfonts,amsmath,times,pict2e}
\usepackage{natbib}
\usepackage{graphicx}
\usepackage{enumerate}
\usepackage{subfigure}
\usepackage{bm}

\usepackage{color}
\usepackage{ulem}
\definecolor{mygray}{gray}{0.6}
\definecolor{magenta}{rgb}{0.858, 0.188, 0.478}

\newcommand{\Rmnum}[1]{\expandafter\@slowromancap\romannumeral #1}

\usepackage{xspace}
\newcommand{\fg}[1]{Fig.~\ref{fig:#1}}
\newcommand{\Fg}[1]{Figure~\ref{fig:#1}}

\newcommand{\eq}[1]{Eq.~(\ref{eq:#1})\xspace}
\newcommand{\Eq}[1]{Equation~(\ref{eq:#1})\xspace}
\newcommand{\eqs}[2]{Eqs.\ (\ref{eq:#1}) and (\ref{eq:#2})}

\newcommand{\se}[1]{Sect.~\ref{sec:#1}\xspace}
\newcommand{\ses}[2]{Sects.\ \ref{sec:#1} and \ref{sec:#2}}

\newcommand{\ie}{i.e.}

\newcommand{\eg}{e.g.}

\newcommand{\AU}{ \  \rm AU}

\newcommand{\Msyr}{ \  \rm M_\odot yr^{-1} }

\newcommand{\vK}{ v_{\rm K}}

\newcommand{\Me}{ \ \rm M_\oplus}

\newcommand{\taus}{ \tau_{\rm s}}

\newcommand{\Omegak}{ \Omega_{\rm K}}

\begin{document}

\title{Growth after the streaming instability}
\subtitle{From planetesimal accretion to pebble accretion}

\author{ Beibei Liu \inst{1,2}, Chris W. Ormel\inst{2}, Anders Johansen \inst{1}}
\institute{
Lund Observatory, Department of Astronomy and Theoretical Physics, Lund University, Box 43, 22100 Lund, Sweden. \label{inst1} 
\and Anton Pannekoek Institute (API), University of Amsterdam, Science Park 904,1090GE Amsterdam, The Netherlands\label{inst2}\\
\email{bbliu@astro.lu.se,c.w.ormel@uva.nl,anders@astro.lu.se}
 }
\date{\today}

\abstract
{
Streaming instability is a key mechanism in planet formation, clustering pebbles into planetesimals with the help of self-gravity. It is triggered at a particular disk location where the local volume density of solids exceeds that of the gas.  After their formation, planetesimals can grow into protoplanets by feeding from other planetesimals in the birth ring as well as by accreting  inwardly drifting pebbles from the outer disk.
}
{
To investigate the growth of planetesimals into protoplanets at a single  location by the streaming instability. For a solar-mass star, we test the conditions under which super-Earths are able to form within the lifetime of the gaseous disk.
}
{
We modify the \texttt{Mercury} N-body code to trace the growth and dynamical evolution of a swarm of planetesimals at a distance of $2.7$ AU from the star. The code simulates gravitational interactions and collisions among planetesimals, gas drag, type I torque, and pebble accretion. Three distributions of planetesimal sizes are investigated: (i) a mono-dispersed population of $400$ km radius planetesimals, (ii) a poly-dispersed populations of planetesimals from $200$ km up to $1000$ km, (iii)  a bimodal distribution with a single runaway body and a swarm of  smaller, $100$ km size planetesimals.
}
{   
 The mono-disperse population of $400$ km size planetesimals cannot form $\gtrsim$ Earth mass protoplanets. Their eccentricities and inclinations are quickly excited, which suppresses both planetesimal accretion and pebble accretion.  Planets can form from the poly-dispersed and bimodal distributions. In these circumstances, it is the two-component nature that damps the random velocity of the large embryo by small planetesimals' dynamical friction, allowing the embryo to accrete pebbles efficiently when it approaches $10^{-2} \Me$. Accounting for migration, close-in super-Earth planets form.  Super-Earth planets are preferred to form when the pebble mass flux is higher, the disk turbulence is lower, or the Stokes number of the pebbles are higher. 
}
{
    For the single site planetesimal formation scenario, a two-component mass  distribution with a large embryo and small planetesimals promotes planet growth, first by planetesimal accretion and then by pebble accretion of the most massive protoplanet. Planetesimal formation at single locations such as ice lines naturally leads to super-Earth planets by the combined mechanisms of planetesimal accretion and pebble accretion.
 }


\maketitle

\section{Introduction} 
\label{sec:intro}

In protoplanetary disks,  micron-sized dust grains  coagulate into pebbles of mm-cm sizes  \citep{Dominik1997,Birnstiel2012,Krijt2016,perez2015,Tazzari2016}. But  further growth is suppressed by bouncing or fragmentation due to the increasing compactification in collisions \citep{Guttler2010,Zsom2010}. In addition, these pebbles also drift too fast compared to their growth \citep{Weidenschilling1977a} so that the particles cannot cross the meter size barrier even if they would stick perfectly \citep{Birnstiel2012,Lambrechts2014}, unless the pebbles can remain fluffy  during the growth \citep{Okuzumi2012,Kataoka2013}. The subsequent growth of these pebbles to planetesimals is still not well understood in planet formation theory (see \citealt{Johansen2014} for a review).

The streaming instability mechanism provides a promising solution by  concentrating drifting pebbles due to a locally enhanced solid-to-gas ratio. Once the threshold of solid-to-gas ratio is satisfied,  the pebble clumps can directly collapse into planetesimals \citep{Youdin2005,Johansen2007,Johansen2009,Bai2010}.
 The characteristic  size of these planetesimals  approximate a few hundred kilometres  \citep{Johansen2012,Johansen2015,Simon2016,Schafer2017,Simon2017,Abod2018}.

The subsequent growth after planetesimal formation by streaming instability has not been well studied. These newly born planetesimals would interact with each other. In the classical planetesimal accretion scenario (see \cite{Raymond2014b,Izidoro2018} for reviews),  the gravitational interactions among these planetesimals lead to orbital crossings, scatterings  and collisions. Initially, the velocity dispersions of planetesimals are not strongly excited and remain modest. In this stage the accretion is super linear ($ {\rm d}m_{\rm p}/{\rm d}t \propto m_{\rm p}^\gamma$ with $\gamma>1$), which is termed `runaway growth' \citep{Greenberg1978,Wetherill1989,Ida1993,Kokubo1996,Rafikov2004,Ida2004a}. It means that the massive body has a faster accretion rate and therefore will get more massive quickly.  However, this stage cannot last forever.  Since the growing massive bodies would stir the random velocities of neighbouring small planetesimals, the accretion rates of the massive bodies slow down and turn into a self-limiting mode. This phase is called   `oligarchic growth' where ${\rm d}m_{\rm p}/{\rm d}t \propto m_{\rm p}^\gamma$ with $\gamma<1$ \citep{Kokubo1998}. This phase is characterized by a decreasing mass ratio of  two adjacent massive runaway bodies \citep{Lissauer1987,Kokubo2000,Thommes2003,Ormel2010b}. 

In addition to planetesimal accretion, planetesimals formed by streaming instability can accrete inward drifting pebbles. A planetesimal  may capture a fraction of the pebbles which cross its orbit \citep{Ormel2010,Lambrechts2012}. This is known as pebble accretion  \citep[see recent reviews by][]{Johansen2017,Ormel2017}. Even if only a fraction of pebbles are able to be accreted by planets, the pebble accretion rate can still be high  for two reasons.  First, the accretion cross section is significantly enhanced  by gas drag \citep{Ormel2010}; and second,  a large flux of pebbles grow and drift inward from the outer regions of disks \citep{Birnstiel2012,Lambrechts2014}. Pebble accretion can be classified into 2D/3D regimes \citep{Ormel2010,Morbidelli2015}. When the pebble accretion radius is larger than the pebble scale height, the accretion is in the $2$D  regime, where ${\rm d}m_{\rm p}/{\rm d}t \propto m_{\rm p}^{2/3}$ \citep[Hill regime]{Lambrechts2014}. On the other hand, when the pebble scale height exceeds the pebble accretion radius, only pebbles with heights smaller than the accretion radius can be accreted. Therefore, the accretion rate in this $3$D regime is reduced compared to $2$D, and ${\rm d} m_{\rm p}/{\rm d}t \propto m_{\rm p}$ \citep{Ida2016}.  

In general, the efficacy of the pebble accretion mechanism to grow planet(esimals) depends on many variables related to the properties of the disk, pebble, and planet (the eccentricity, the inclination, and the mass of the planet, the pebble size, the disk turbulence, etc.)
A key quantity is the pebble accretion efficiency $\varepsilon_{\rm PA}$ \citep{Guillot2014,Lambrechts2014}, defined as the number of pebbles that are accreted divided by the total number of pebbles that the disk supplies. 
Recently we have computed  $\varepsilon_{\rm PA}$  under general circumstances \citep{Liu2018,Ormel2018}.
For instance, when the eccentricity and inclination of the planet become high, $\varepsilon_{\rm PA}$ drops significantly compared to planets on coplanar and circular orbits, because pebbles are approaching at too high velocity to the planet.

Progress in planet formation  requires an improved understanding  under which conditions the mass growth is dominated by accreting pebbles  or planetesimals. In this work, our goal is to investigate the growth of planetesimals after their formation by streaming instability at a single disk location (\eg, the H$_2$O ice line).  \cite{Hansen2009} already proposed that the architecture of the Solar system's terrestrial planets can be explained when planetesimals grow in a narrow annulus.
In his model the width of the annulus is $0.3$ AU, much wider than our planetesimal forming zone (see \se{method}). Furthermore, \cite{Hansen2009} focused on the planetesimal accretion in a gas-free environment. Our work instead considers the growth just after the streaming instability in gas-rich disk phase.

Constrained by the size distribution of planetesimals in the asteroid belt, \cite{Morbidelli2009} concluded that their born size should be large ($\gtrsim 100$ km), while \cite{Weidenschilling2011} argued that planetesimals starting from sub-km-sized still cannot be ruled out. 
\cite{Kenyon2010} studied the formation of ice planets beyond $30$ AU and found that protoplanets grow more efficiently with smaller planetesimal sizes.  
Motivated by streaming instability simulations, our adopted initial sizes are typical $\gtrsim 100 $km. In the context of combined planetesimal and pebble accretion,  \cite{Johansen2015} studied the growth of asteroids using a statistical approach, and concluded that massive protoplanets or even super-Earths can form by a combination of pebble accretion, planetesimal accretion and giant impacts.

In order to study planet formation from a narrow ring of planetesimals, we employ direct N-body techniques. Growth can be classified into two phases: (A) the planetesimal accretion dominated phase and (B) the pebble accretion dominated phase.  The N-body approach is necessary to treat phase A and the transition from phase A  to phase B. The \texttt{Mercury} N-body code has been modified  to include gas drag, type I torque and pebble accretion.  Three different types of initial planetesimal size distributions are investigated. We find that a two-component mass distribution (large embryo + small planetesimals) is needed to grow a massive planet. This condition could arise either from the high mass tail distribution of planetesimals formed by streaming instability or be the result of runaway growth of a population of small planetesimals.

The paper is structured as follows. In \se{method}, we outline our model and the implementation of the N-body code.  
In \se{threesizes} three initial size distributions of planetesimals are investigated, including a mono-dispersed population in \se{mono-dispersed}, a poly-dispersed population in \se{poly-dispersed}, and a single  runaway body plus a swarm of small planetesimals in \se{two-component}. In \se{parameter} we  investigate the influence of different  parameters in the pebble accretion dominated growth phase (phase B). The  key results are summarized  in \se{summary}.

\section{Method}
\label{sec:method}
The hypothesis of this paper is that planetesimals only form at a specific location by streaming instability, which requires a locally enhanced pebble density \citep{Carrera2015,Yang2014,Yang2017,Yang2018}.  For instance,  \cite{Ros2013,Schoonenberg2017,Drazkowska2017} have proposed that the ice line could be such a place since the water vapor inside the ice line will diffuse back and re-condense onto ice pebbles, enriching the solid to gas density ratio. We focus on planetesimals that form at the ice line ($r_{\rm ice} =2.7  \ \rm AU$ based on the disk model in \se{disk}) in this paper. But the following results and applications  can be scaled to other locations where the streaming instability condition is realized, \eg , the inner edge of the  zone \citep{Chatterjee2014,Hu2018} or a distant location due to the FUV photoevaporation \citep{Carrera2017}. In order to generalize our results, we did not include the specific ice line effects such as a reduced pebble size and pebble flux inside of the ice line due to sublimation.

\begin{figure}[t]
        \includegraphics[scale=0.19, angle=0]{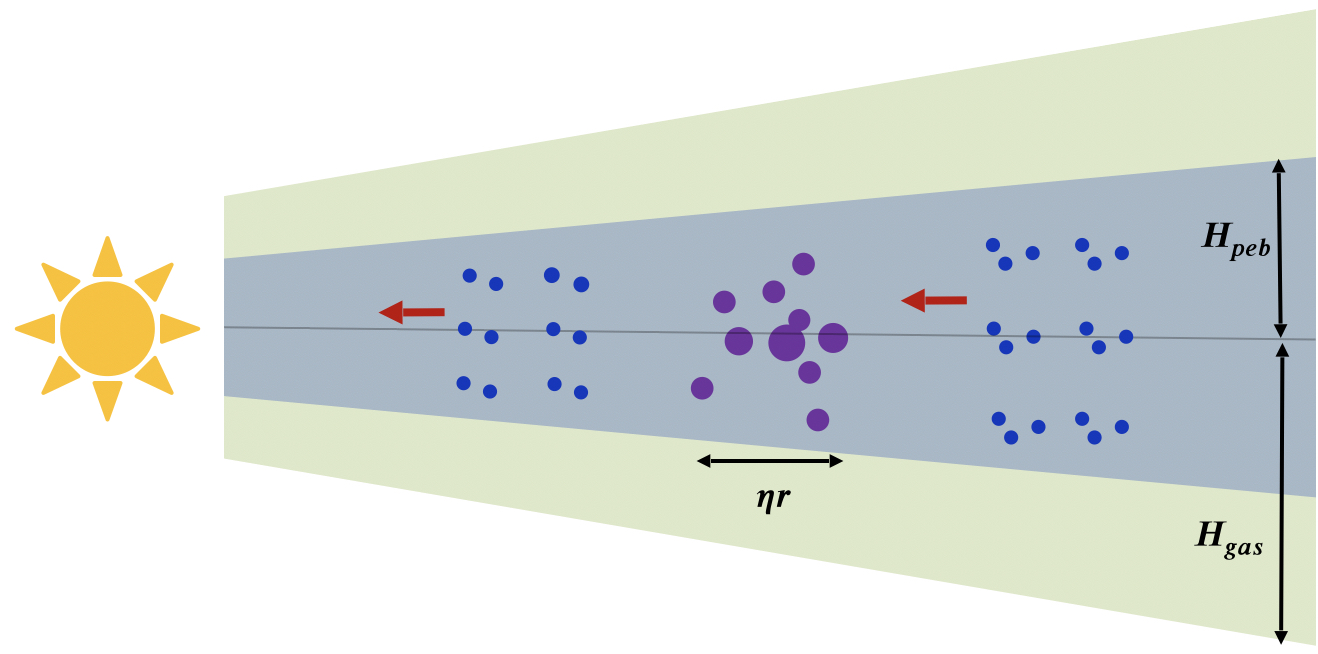}
       \caption{Sketch illustrating the initial setup. Planetesimals (purple circles) have formed at a single location by streaming instability. They accrete both among themselves as well as from the inward drifting pebbles (blue). The pebble disk and gas disk are marked in light blue and light green, where the gas and pebble scale heights are $H_{\rm gas}$, $H_{\rm peb}$, respectively. The horizontal black arrow represents the radial width ($\eta r$) of the forming planetesimal ring. Due to dynamical friction, large planetesimals have low eccentricities and inclinations whereas small planetesimals have high velocity dispersions.
      }
\label{fig:carton}
\end{figure} 

This scenario is illustrated in \Fg {carton}.
We consider the situation where  streaming instability operates  to quickly spawn planetesimals at the initial time of our simulations.  A population of planetesimals has formed in a narrow annulus of relative width equal to the normalized pressure gradient ($\Delta r = \eta r$, see \fg {carton} and further discussion in \se{planetesimal initial condition}). These planetesimals can grow further in two ways: by coagulation among themselves (planetesimal accretion) or by sweeping up pebbles that drift in from the outer disk (pebble accretion). 

\subsection{Disk model}
\label{sec:disk}
The surface density and the aspect ratio of the gas disk are assumed to be   
 \begin{equation}
  \Sigma_{\rm gas} =  \Sigma_{\rm gas0} \left(  \frac{r}{1 \AU} \right)^{-1}, 
  \label{eq:siggas}
\end{equation}   
 \begin{equation}
  h_{\rm gas} =  h_{\rm gas0} \left(  \frac{r}{1 \AU} \right)^{1/4}, 
  \label{eq:hgas}
\end{equation}
where $ \Sigma_{\rm gas0}$ and $h_{\rm gas0}$ are the gas surface density and the aspect ratio at $1$ AU and $r$ is the distance to the central star. 
The aspect ratio index of 1/4 assumes an optically thin stellar irradiated disk. For simplicity, we neglect that in the early stages disks might be hotter due to viscous accretion \citep{Garaud2007,Bitsch2015}.
Here $ \Sigma_{\rm gas0} = 500 \rm \ g\,cm^{-2}$ and $h_{\rm gas0} = 0.033$  \citep{Hayashi1981} are adopted as the default values in this paper. 
Therefore, the disk temperature $T$ and  $\eta$ are 
 \begin{equation}
  T =  \frac{ \mu h_{\rm gas}^2  G M_{\star}}{r R_{\rm g}}  = T_0 \left(  \frac{r}{1 \AU} \right)^{-1/2}, 
  \label{eq:temp}
\end{equation}
and 
 \begin{equation}
  \eta =  -\frac{h_{\rm gas}^2}{2}  \frac{\partial {\rm log} P_{\rm gas} }{ \partial {\rm log} r} = \eta_0 \left(  \frac{r}{1 \AU} \right)^{1/2}, 
  \label{eq:eta}
\end{equation}
where $G$ is the gravitational constant, $R_{\rm g}$ $=$ $8.31 \times10^{7} \rm erg \, mol^{-1} \, K^{-1} $ is the gas constant, $M_{\star}$ is the stellar mass, $\mu=2.34$ is the mean molecular weight and $P_{\rm gas}$ is the gas pressure in the disk.  In the above equations  $T_{0} = 280 \rm \ K $ and  $\eta_0=1.5 \times 10^{-3}$ are the values at $1$ AU.

 Although the adopted surface density power-law index is shallower than minimum mass solar nebula model, this profile ($\Sigma_{\rm g} \propto r^{-1}$) is more consistent with disk observations \citep{Andrews2009}. The fiducial value of $\Sigma_{\rm gas0} $ is chosen to be the same as \cite{Lambrechts2014}, and the pebble flux is calibrated accordingly in \se{pebble}. 
Based on the above disk model, the ice line ($T\sim 170 \rm K$) is located at  $r_{\rm ice} = 2.7$ AU.

\subsection{Simulation setup}
We use numerical N-body simulations to study the planetesimal growth after the streaming instability during the gas-rich disk phase.  
We have adopted the Mercury code \citep{Chambers1999} and used the Bulirsch-Stoer integrator.  An initial timestep of $3$ days and the integration accuracy parameter of $10^{-12}$ are chosen.

 For the N-body part, collisions between bodies are treated as inelastic mergers that conserve the linear momentum.
 The fragmentation/restitution \citep{Leinhardt2012,Mustill2018} is not taken into account in this work. Since eccentricities are initially low, the perfect merger assumption is appropriate. For instance, the impact velocity of $400$-km-sized planetesimals is lower than the escape velocity when their eccentricities are lower than a few times $10^{-2}$.  Fragmentation among planetesimals will become important after embryos form and stir the planetesimals more vigorously. However, by then pebble accretion is already expected to commence, rendering planetesimal fragmentation irrelevant.

The effects of the disk gas on the planetesimals/planets, such as gas drag, type I torque, eccentricity and inclination damping are taken into account by applying effective forces in \se{gas}.  In \se{pebble}, we implement the pebble accretion prescriptions by \cite{Liu2018} and \cite{Ormel2018}, which accounts for the effects of the planet's eccentricity, inclination and the disk turbulence.   The modified code uniquely simulates planet-planet, planet-disk and planet-pebble interactions.

\subsubsection{Gas drag and type I migration torque}
\label{sec:gas}
Small embryos and planetesimals experience  the aerodynamic gas drag \citep{Adachi1976}, 
\begin{equation}
  \bm{a_{\rm drag}} =  - \left( \frac{ 3 C_{\rm D} \rho_{\rm gas} }{8  R_{\rm p}  \rho_{\bullet}  } \right) v_{\rm rel} \bm{v_{\rm rel}}, 
  \label{eq:adrag}
\end{equation}
where the drag coefficient $C_{\rm D} =0.5$, $v_{\rm rel}$  is the relative velocity between the planetesimal and the gas, $\bm{v}_\mathrm{rel} = \bm{v}-\bm{v}_\mathrm{gas}$, where $\bm{v}_\mathrm{gas}$ equals $ v_{\rm K}(1-\eta)$ in the azimuthal direction, $\vK$ is the Keplerian velocity, $\rho_{\rm gas}$ is the local gas density, $\rho_{\bullet}$ and   $R_{\rm p}$ are the internal density (assumed to be $1.5 \rm \ g\, cm^{-3}$, with half water and half silicate rock) and the physical radius of the planetesimal.   

 Large embryos and low-mass planets feel the gravitational torques from the disk gas (called type I migration, \cite{Goldreich1979,Kley2012,Baruteau2014}). The characteristic migration timescale  for a planet on a circular orbit is \citep{Cresswell2008}
 \begin{equation}
 \begin{split}
 t_{\rm m}  \simeq &  0.5 \left( \frac{M_{\star}}{ m_{\rm p}} \right)  \left( \frac{M_{\star}}{ \Sigma_{\rm gas} a_{\rm p}^2  } \right)  \left( \frac{ h_{\rm gas}^2 }{\Omega_{\rm K}} \right)
  \simeq 5\times10^{5} \left( \frac{m_{\rm p}}{ 1 \Me } \right)^{-1}  \\
  &\left( \frac{\Sigma_{\rm gas0}}{500 \rm \ g cm^{-1}} \right)^{-1}
 \left( \frac{h_{\rm gas0}}{3.3 \times 10^{-2}} \right)^{2} 
  \left( \frac{a_{\rm p} }{2.7  \AU} \right)  \rm \ yr,
  \label{eq:twave}
  \end{split}  
\end{equation}
where $m_{\rm p}$ and $a_{\rm p}$ are the mass and the semimajor axis of the planet, respectively, and $\Omega_{\rm K}$ is the Keplerian angular velocity at the planet location.

The accelerations acting on planets due to type I migration torque are 
   \begin{equation}
 \bm {a_{\rm m}} = -\frac{\bm{v}}{t_{\rm m}}, \ \bm {a_{\rm e}} = -2 \frac{(\bm{v} \cdot \bm{r}) \bm{r}}{ r^2 t_{\rm e}}, \  \bm {a_{\rm i}} = - \frac{\bm{v_{\rm z}}}{ t_{\rm i}},
  \label{eq:amig}
 \end{equation}
 where $\bm{r} = (x,y,z)$, $\bm{v} = (v_{\rm x},v_{\rm y},v_{\rm z})$  are the position and the velocity vectors of the planet. In the above expression, $t_{\rm m}$, $t_{\rm e}$ and $t_{\rm i}$ are the characteristic type I migration, eccentricity and inclination damping timescales from Eqs. (13), (11) and (12) of \cite{Cresswell2008}. We note that the torque prescription is based on an isothermal disk and planets always migrate inward. The effect of the unsaturated corotation torque due to the thermal diffusion in a radiative disk \citep{Paardekooper2010,Paardekooper2011,Bitsch2015,Brasser2017}, the dynamical torque \citep{paardekooper2014,McNally2018}, and the heating torque from planet gas accretion \citep{Benitez-Llambay2015,Masset2017,Chrenko2017} are not taken into account in this work.
 The mentioned migration prescriptions are implemented in all our model runs except for \textit{run\_sp\_nmig}  in \se{parameter} (see Table 2). In that case we neglect the semimajor axis damping  (set $\bm {a_{\rm m} =0}$) but still consider the eccentricity and inclination damping in \eq{amig} due to the type I torque.  This setup would mimic the situation when the planet is situated at a net zero-torque location in the disk where the negative Lindblad torque and positive corotation torque are cancelled out.

\subsubsection{Pebble accretion} 
\label{sec:pebble}
The pebble-sized particles drift inwards across the protoplanetary disk. The radial drift velocity is $v_{\rm r} = -2 \eta v_{\rm K} \tau_{\rm s}/(1 + \tau_{\rm s}^2)$ \citep{Weidenschilling1977a} where $\taus$ is the dimensionless stopping time (Stokes number). The drift speed is determined by the aerodynamical size of the pebble $\taus$ and  $\eta$.  Based on \cite{Lambrechts2014} and \cite{Schoonenberg2018a}, 
the pebble mass flux ($\dot M_{\rm peb} = 2\pi r  v_{\rm r}\Sigma_{\rm peb} $)
is proportional to the disk pebble surface density, but weakly dependent on time. In the default model we adopt a constant pebble flux of $\dot M_{\rm peb}= 100 \ \rm M_{\oplus}/Myr $ (consistent  with Eq. (14) of \cite{Lambrechts2014}) and neglect the time dependence for simplicity. A lower $\dot M_{\rm peb}$ means an intrinsic metal-deficient/less massive disk and the planetesimals/planets take longer time to grow by pebble accretion.  On the other hand, the pebble flux cannot be too high as the density ratio between pebbles and gas would otherwise exceed unity ($\rho_{\rm peb}/\rho_{\rm gas} >1$) and the streaming instability will be triggered over the entirely disk. The above 
inequality can be written as   $\Sigma_{\rm peb}/H_{\rm peb} > \Sigma_{\rm gas}/H_{\rm gas}$ where the scale height of the pebble is given by \citep{Youdin2007}
 \begin{equation}
 H_{\rm peb} = \sqrt{\frac{\alpha_{\rm t}}{\alpha_{\rm t} + \taus}} H_{\rm gas}.
  \label{eq:Hpeb}
\end{equation}  
The pebble flux therefore needs to be smaller than $2\pi r  v_{\rm r}\Sigma_{\rm gas} H_{\rm peb}/H_{\rm gas}$. Our default value is well below this limit. 
We also explore two additional pebble fluxes    $\dot M_{\rm peb} = 200 \Me Myr^{-1}$ and $50 \Me Myr^{-1}$  in \se{parameter}.

The pebble mass can be measured from (sub)millimeter  dust continuum emission  from the young protoplanetary disks. The inferred values are from a few Earth mass to a few hundreds of Earth mass \citep{Ricci2010,Andrews2013,Ansdell2017}, which is correlated with the gas disk accretion rates \citep{Manara2016}.  For a typical T Taurs star with a disk accretion rate of $10^{-8} \ \rm M_{\odot}\, yr^{-1}$, the dust mass is $\sim 100 \Me$ (Fig.1 of \cite{Manara2016}), consistent with our adopted fiducial value.

 In \eq{Hpeb} $\alpha_{\rm t}$ is the coefficient of turbulent gas diffusivity, which is different from the concept of turbulent viscosity $\alpha_{\nu}$.  We nevertheless note that the above two quantities are approximated the same when the turbulence is driven by magnetorotational instability \citep{Johansen2005,Zhu2015,Yang2018}.  The value of $\alpha_{\rm t}$ can be constrained from the molecule line broadening measurements \citep{Flaherty2015,Flaherty2017} and the level of dust settling \citep{Pinte2016}.
We adopt a  fiducial value of $\alpha_{\rm t}  =10^{-3}$ and test the influence of a lower turbulent disk  ($\alpha_{\rm t}  =10^{-4}$) in \se{parameter}.

For pebbels we adopt a fiducial aerodynamical size of $\taus =0.1$. This chosen value is consistent with the sophisticated dust coagulation model \citep{Birnstiel2012} and disk observations \citep{Tazzari2016}.  A lower $\taus =0.03$ is also explored in \se{parameter}.

A fraction of pebbles will be accreted onto a planetesimal when pebbles drift through its orbit. The pebble accretion efficiency ($\varepsilon_{\rm PA} = \dot M_{\rm PA}/ \dot M_{\rm peb}$) is taken from  \cite{Liu2018} and \cite{Ormel2018}. 
This quantity depends on the disk properties ($\taus$, $\eta$, $\alpha_{\rm t}$) and the planet properties ($m_{\rm p}, a, e, i$).  In the limit of 2D and 3D pebble accretion these are given by
\begin{equation}
    \label{eq:eps-2D}
    \varepsilon_\mathrm{2D} = \frac{0.32}{\eta} \sqrt{\frac{m_{\rm p} \Delta v}{M_\star v_{\rm K} \tau_s}} f_\mathrm{set}
\end{equation}
and
\begin{equation}
    \label{eq:eps-3D}
    \varepsilon_\mathrm{3D} = 0.39 \frac{m_{\rm p}}{\eta h_\mathrm{peb} M_\star} f_\mathrm{set}^2
\end{equation}
respectively, where $\Delta v$ is the relative velocity between the planet and the pebble, $h_\mathrm{peb}= H_{\rm peb}/r$ is the aspect ratio of the pebble disk.  In \eq{eps-3D} we have already assume that $i< h_{\rm peb}$. The above expressions include a modulation factor,  
\begin{equation}
    \label{eq:fset0}
    f_\mathrm{set} = \exp\left[ -0.5 (\Delta v/v_\ast)^2 \right]
\end{equation}
where $v_\ast=(M_p/\tau_s M_\star)^{1/3} v_{\rm K}$ \footnote{A more general form of $f_{\rm set}$ can be found in Eq. (35) of \cite{Ormel2018}, considering the pebble's velocity distribution in a three-dimensional turbulent disk.}. Physically, when the pebble-planet encounter is too fast compared to the coupling time between the gas and the pebble, gas drag is no longer effective to aid the planet to capture pebbles. Therefore, $f_{\rm set} \ll1$ and pebble accretion fails \citep{Visser2016}.

In the multi-planetesimal system we consider the filtering of  flux when pebbles drift through these planetesimals. Therefore, the pebble flux of the body $i$ is given by 
\begin{equation}
    \dot M_{\rm i,peb} = \left \{
    \begin{aligned}
 & \dot M_{\rm peb} &  i = 1 \\
 & \dot M_{\rm peb} \prod_{k=1}^{i-1}(1-\varepsilon_{\rm k,PA} ),  &    i\geq 2 \\
    \end{aligned}
    \right.
    \label{eq:mdot}
\end{equation} 
where $i$ is ordered for bodies from the furthest to closest in terms of semimajor axis and $\varepsilon_{i,\mathrm{PA}}$ is the pebble accretion efficiency of the $i^\mathrm{th}$ body.
  
We  focus on the formation of protoplanets with $m_{\rm p } \gtrsim 1 \Me$ (progenitors of super-Earths). In our simulations the forming  planets are still not massive enough to inverse the local gas pressure gradient and truncate the inward drift of pebbles \citep{Lambrechts2014b,Bitsch2018,Ataiee2018}. Therefore, we do not implement  the termination of pebble accretion.

\subsection{Planetesimal initial condition}
\label{sec:planetesimal initial condition}

In the streaming instability mechanism, pebbles accumulate into dense filaments. The typical width of the filament  is $\Delta r \simeq \eta r_{\rm ice} $ \citep{Yang2016,Li2018}.  The threshold condition to trigger gravitational collapse requires $\rho_{\rm peb} \simeq \rho_{\rm gas}$.  The total solid mass available to build planetesimals is therefore $2\pi r_{\rm ice} \Delta r  \Sigma_{\rm peb}(r_{\rm ice})  =2\pi r_{\rm ice} \Delta r  \Sigma_{\rm gas}(r_{\rm ice}) H_{\rm peb}/H_{\rm gas} $.  \cite{Simon2016} find  that the planetesimal generation efficiency  approximates $50\%$ (half of the pebbles convert into planetesimals, see their Fig. 10a). We note that this value might be also dependent on the local metallicity, Stokes number of the pebbles, and the disk turbulence. However, due to a lack of  detailed streaming instability simulations exploration,  we still use this number to approximate the total mass in planetesimals.  With our adopted disk parameters, this amounts to $0.039 \ M_{\oplus}$.

\section{Scenarios with different initial sizes}
\label{sec:threesizes}

 \begin{figure*}[t!]
    \sidecaption    
      \includegraphics[scale=0.49, angle=0]{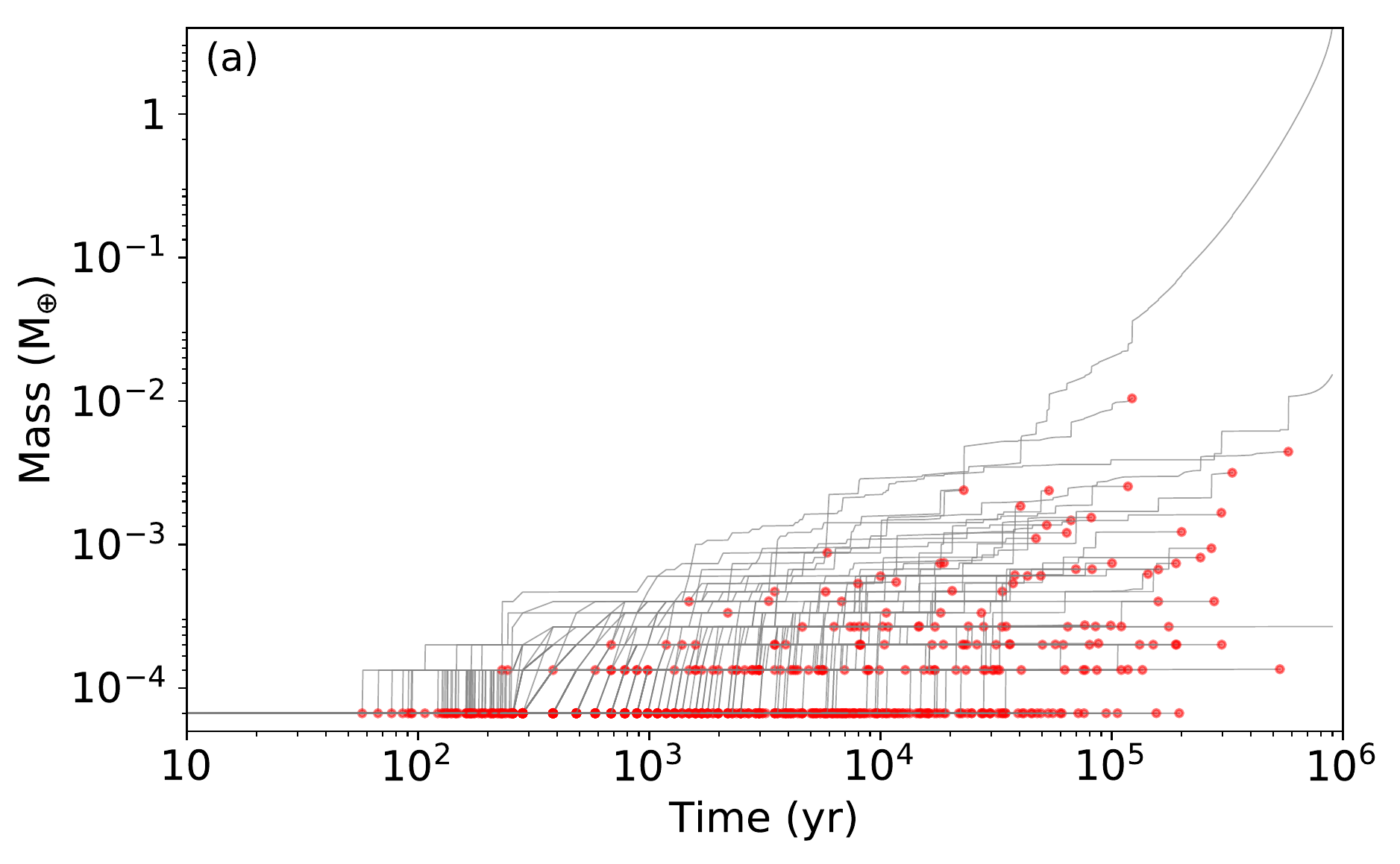}
       \includegraphics[scale=0.49, angle=0]{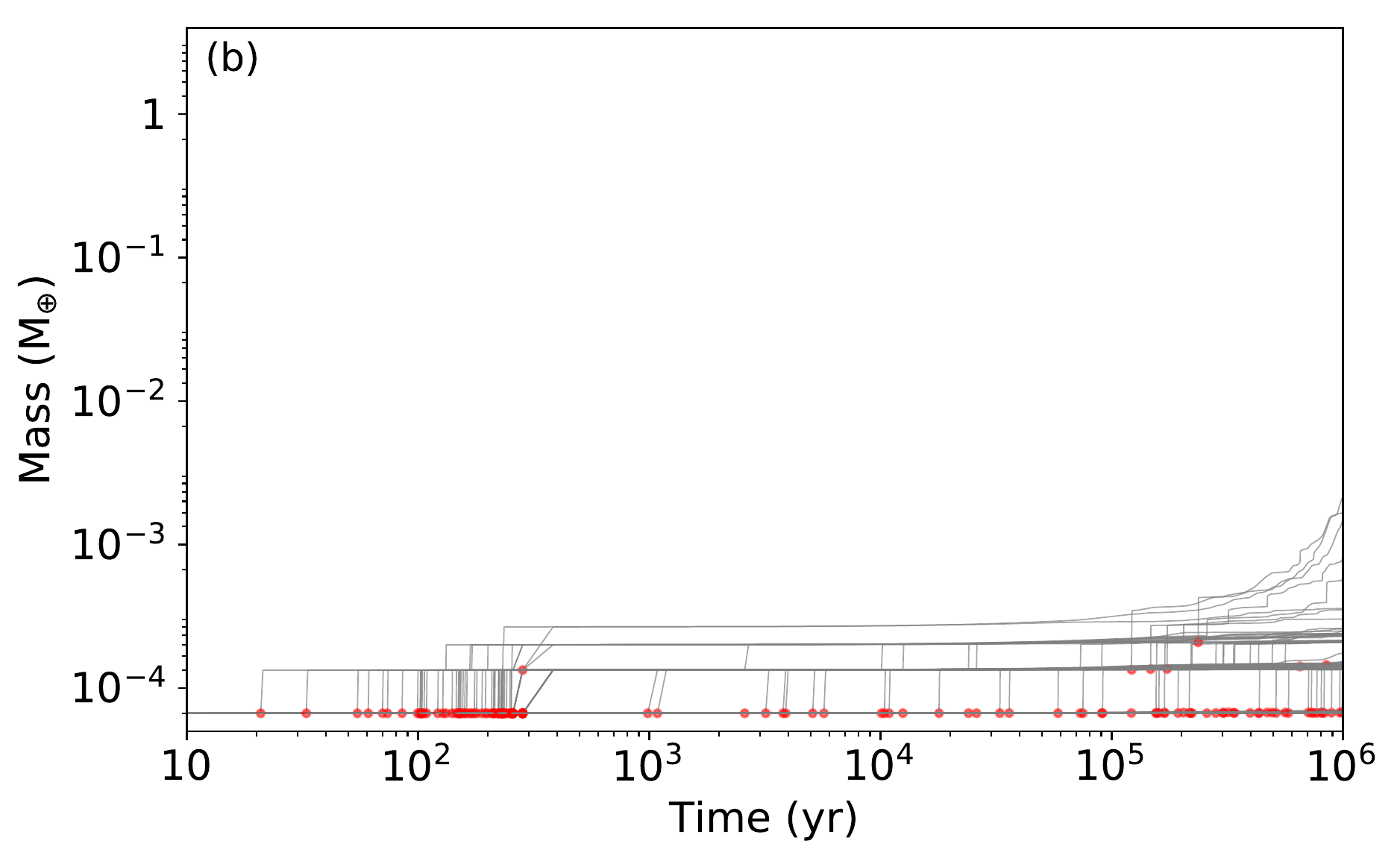}
       \caption{Mass  growth of planetesimals for  planets on coplanar orbits (left, \textit{run\_md\_test} in Table 1) and non-coplanar orbits (right, \textit{run\_md\_1}). Grey lines trace the mass of each planetesimal. Red dots represents the merging of two bodies, marked at the position of the less massive body.   The early sudden growth is caused by planetesimal accretion and the later smooth growth by pebble accretion. 
    }
\label{fig:2d3d}
\end{figure*}

Many works have simulated streaming instability numerically, finding that it spawns planetesimals of typical size of several hundreds kilometers, albeit with a considerable uncertainty regarding the precise shape of the size distribution \citep{Johansen2015,Simon2016,Schafer2017,Abod2018}. However, this initial distribution affects the planet growth, as it will determine the duration of the planetesimal-dominated growth phase (phase A). Here we consider three scenarios for the initial planetesimal distribution in the following subsections: (i) a mono-dispersed population of big planetesimals; (ii) a population of planetesimals of various sizes; and (iii) a two-component population of one large body among many small planetesimals. The first and third scenario reflects standard practice where planetesimals typically start from a fixed size, whereas the second more closely follows the streaming instability results. While the first and second case can be integrated directly, the third scenario employs a superparticle approach to handle the large number of small bodies with N-body techniques.

In this section we investigate the question that provided in total $100 \Me$ of pebbles ( $100\Msyr$ of $\dot M_{\rm peb}$  over $1$ Myr), what is the mass growth of planetesimals for the above three different initial size distributions.

\subsection{Mono-dispersed initial conditions}
\label{sec:mono-dispersed}
In this scenario we consider that all planetesimals start with an equal size ($400$ km in radius).
The mass growth and dynamical evolution are simulated after the formation at a single site location. We conduct both  simulations when the planetesimals orbits are coplanar and inclined. In the former case planetesimals are all initiated at $z=0$ and, by construction, remain in the disk midplane.  This (as we will see) is not realistic, but comparing the results between these two gives us important insights on key factors (\eg , planet inclination) shaping the planetesimal + pebble accretion  process.

\begin{table*}
    \centering
    \caption{Simulations set-up in \se{mono-dispersed}, \se{poly-dispersed} and \se{two-component}. The mass and semi-major axis of the most massive planet are given at $t =1$ Myr.}
    \begin{tabular}{lclclclclclc|}
        \hline
        \hline
        Name    &  initial planetesimal size  & inclination   &  $ m_{\rm p} \ (M_{\oplus})$ &  $a_{\rm p}$  (AU)  \\ 
      \hline
      \textit{run\_md\_test} & mono-size &No  & $4.0$ &$0.1$ \\
      \textit{run\_md\_1} & mono-size &Yes  & $0.004$ &$2.7$ \\
      \textit{run\_md\_2} & mono-size &Yes  & $0.009$ &$2.7$ \\
      \textit{run\_md\_3} & mono-size &Yes  & $0.02$ &$2.6$ \\
      \textit{run\_pd\_1} & poly-size &Yes   & $1.3$ &$1.7$ \\
      \textit{run\_pd\_2} & poly-size &Yes   & $1.6$ &$1.5$ \\
      \textit{run\_pd\_3} & poly-size &Yes   & $1.4$ &$1.8$ \\
      \textit{run\_sp\_1} & one embryo + small planetesimals   &Yes   & $1.3$ &$1.8$ \\
      \textit{run\_sp\_2} & one embryo + small planetesimals    &Yes   & $1.4$ &$1.6$ \\
      \textit{run\_sp\_3} & one embryo + small planetesimals    &Yes   & $1.3$ &$1.7$ \\
          \hline
        \hline
    \end{tabular}
    \label{tab:tab1}
\end{table*}

\subsubsection{Effect of inclinations on  planetesimals growth}
\label{sec:2Dvs3D}
As mentioned in \se{planetesimal initial condition},  the streaming instability  produces in total $0.039 \ M_{\oplus}$ planetesimals at $r_{\rm ice}$. In the mono-dispersed initial conditions,  we assume that all  planetesimals are $400 \rm \ km$ in radius. ($6.7 \times 10^{-5} \Me$ in mass). Therefore, $N=580$  planetesimals are generated in this compact ring belt ($\Delta r = \eta r$ in width). These bodies are injected into the simulation one by one after each  $0.1$ orbital period. Their initial semimajor axes are uniformly distributed from $r_{\rm ice}-\Delta r /2$ to $r_{\rm ice} + \Delta r/2$.   The initial eccentricities are assumed to follow a Rayleigh distribution,  $p(e) =  e/e_0 \exp [- e^2/2e_0^2]$.

We conduct two sets of simulations. The first  idealised set considers that planetesimals are in coplanar orbits ($i_{\rm p}= 0 $, $run\_md\_test$ in Table 1). However, this configuration ($i_0 =0$) is an  unphysical case. Realistically, although the initial inclinations are tiny, they are not zero due to the stochastic fluctuation driven by the disk turbulence \citep{Ida2008,Gressel2011,Yang2012,Okuzumi2013}. The planetesimals would be lifted out of the midplane and acquire an inclination distribution.

In hydrodynamic simulations \citep{Simon2016,Schafer2017} the planetesimals are generated from pebble filaments, their random fluctuation is at least smaller compared to the size of the filament ($e_0$, $i_0 \ll \eta $).  Therefore, for the second set we assume that  their inclinations also follow a Rayleigh distribution with $i_0= e_0/2$.  Two different initial values for $e_0$ are numerically tested, $10^{-5}$ and $10^{-6}$.  We find that since the orbit of planetesimals are readily excited, the results are insensitive to the initial values.  
In this non-coplanar configuration, three different simulations are performed with the randomlized  initial semi-major axes and orbital phase angles  (\textit{run\_md\_1} to \textit{run\_md\_3} in Table 1).

\Fg{2d3d} shows the mass growth of planetesimals when they are in co-planar orbits  (panel a) and inclined orbits  (panel b).  We clearly find that mergers (red dots) occur more frequently  when the orbits of the planetesimals are coplanar. In \fg{2d3d}a the mass growth is dominated by planetesimal-planetesmial collision at the beginning (the sudden jump in mass of grey lines in \fg{2d3d}). Since the pebble accretion rate is an increasing function of the planet mass,  when the mass approaches  $ 10^{-2} \ M_{\oplus}$, the growth is driven by pebble accretion (smooth growth in mass  in \fg{2d3d}). Only a few bodies survive after $1$ Myr, and the mass of the dominant body is  $ 4 \ M_{\oplus}$. Even though the accretion timescale in this configuration is artificially short, it clearly illustrates that the growth initially proceeds in a planetesimal accretion-dominated phase (phase A) and transitions to a rapid pebble accretion-dominated phase (phase B) when a massive body of $\sim$$10^{-3} \Me$ to $ 10^{-2} \Me$ emerges. 
For the realistic case when planetesimals are on inclined orbits  (\fg{2d3d}b),  however, much fewer mergers occur and the mass growth remains modest. The mass of the largest body remains $ < 10^{-2} \Me$, and there is no efficient pebble accretion at the end of the simulation.

\subsubsection{Excitation of eccentricities and inclinations}
\label{sec:excitation}
\fg{eccinc} shows the time evolution of the root-mean-square (rms) of the planetesimals' eccentricities (orange) and inclinations (cyan) in  \textit{run\_md\_1}. After a few orbits the eccentricities are quickly excited while the inclinations still remain low. This behavior  occurs in the shear-dominated regime, when the velocity dispersion ($\delta v$) of planetesimals is smaller than their Hill velocity ($v_{\rm H} =R_{\rm H} \Omegak$) \citep{Ida1990}, where the mutual Hill radius  $R_{\rm H} = (2m_{\rm p}/3M_{\star})^{1/3} a$.  After a few hundred years their eccentricities get excited to values larger than the Hill velocity ($\delta v > v_{\rm H} $, dashed line in \fg{eccinc}), the scattering transitions to the isotropic, dispersion-dominated regime. In that case the rms inclination excites to a value equals to half of the eccentricity, $i_{\rm rms} \simeq e_{\rm rms}/2$ \citep{1993b}.

Based on the disk model, we calculate the excitation and damping timescale of the velocity dispersion of the planetesimals.   
The timescale of the viscous stirring (eccentricity excitation) is given by \citep{Ida1990,Kokubo2000},
 \begin{equation}
  \begin{split}
     & \tau_{\rm  vs}   \simeq   3 \times 10^{5}  \left( \frac{i}{0.005} \right) \left[  \left(\frac{e}{0.01} \right)^2 + \left( \frac{i}{0.005} \right)^2  \right]^{\frac{3}{2}}  \left(  \frac{R_{\rm p}}{400 \ \rm km} \right)^{-3}   \\
   &   \left(  \frac{\Sigma_{\rm p}}{10 \rm \ g\,cm^{-2}}  \right)^{-1}   \left(  \frac{\rho_{\bullet}}{1.5 \rm \ g\,cm^{-3}}  \right)^{-1} \left(  \frac{a}{2.7 \AU}  \right)^{-1/2} \  \rm yr.
   \end{split}
   \label{eq:tau_vs}
\end{equation}
The damping timescale of eccentricity from the gas drag is \citep{Adachi1976},
 \begin{equation}
  \begin{split}
  \tau_{\rm  drag}  \simeq  &   10^{7}  \left(  \frac{R_{\rm p}}{400 \ \rm km} \right) \left(  \frac{\Sigma_{\rm gas0}}{500 \rm \ g\,cm^{-2}}  \right)^{-1} \left(  \frac{h_{\rm gas0}}{0.033}  \right)  \\ 
  & \left[ \left( \frac{e}{10^{-2}} \right)^2 +\left( \frac{i}{5 \times 10^{-3}} \right)^2 +  \left( \frac{\eta}{2.5 \times 10^{-3}} \right)^2 \right]^{-1/2}  \\ 
  &     \left(  \frac{\rho_{\bullet}}{1.5 \rm \ g\,cm^{-3}}  \right) \left(  \frac{a}{2.7 \AU}  \right)^{11/4} \ \rm yr.
   \end{split}
   \label{eq:tau_drag}
\end{equation}
This damping is inefficient for large planetesimals, and is a strongly increasing function of  the semi-major axis of  the planet $a$. 
Tidal interactions (type I torque) between an embryo and  the  gas  also damps its eccentricity and inclination. The  damping terms is expressed as \citep{Artymowicz1993} 
 \begin{equation}
  \begin{split}
&  \tau_{\rm tidal}    \simeq   
  \left( \frac{M_{\star}}{ m_{\rm p}} \right)  \left( \frac{M_{\star}}{ \Sigma_{\rm gas} a_{\rm p}^2  } \right)    \left( \frac{ h_{\rm gas}^4 }{\Omega_{\rm K}} \right)
   \simeq 7 \times 10^{7}  \left(  \frac{R_{\rm p}}{400 \ \rm km} \right)^{-3}  \\ 
   & \left(  \frac{\Sigma_{\rm gas0}}{500 \rm \ g\,cm^{-2}}  \right)^{-1} 
   \left(  \frac{h_{\rm gas0}}{0.033}  \right)^4    \left(  \frac{\rho_{\bullet}}{1.5 \rm \ g\,cm^{-3}}  \right)^{-1}  \left(  \frac{a}{2.7 \AU}  \right)^{3/2} \ \rm yr.
   \end{split}
   \label{eq:tau_tidal}
\end{equation}
The type I damping is inversely proportional to the mass of the embryo, and is negligible when the size of the body is smaller  than $1000$ km ($10^{-3} \Me$). 
From the above equations we obtain $\tau_{\rm vs} \ll \tau_{\rm drag}  <  \tau_{\rm tidal}$. 
We conclude that both  gas drag and type I damping are ineffective  to circularize the orbit of planetesimals. Their eccentricities and inclinations  keep increasing with time.

The next question is why pebble accretion is inefficient in the simulation in \fg{2d3d}b? There are two reasons. First, fewer planetesimal collisions result in less massive planetesimals. Since the pebble accretion rate correlates with the mass of the body (\eqs{eps-2D}{eps-3D}), less massive bodies accrete pebbles more slowly.  Second, pebble accretion is quenched due to a high eccentricity and inclination. Specifically, from \eq{fset0} it follows that pebble accretion requires encounters to have a sufficiently low relative velocity
\begin{equation}
    \label{eq:PA-condition}
    \Delta v \lesssim v_\ast = \left( \frac{M_p/M_\star}{\tau_{\rm s} } \right)^{1/3} v_{\rm K}.
\end{equation}
Inserting $\Delta v \simeq ev_{\rm K}$ and $e \sim 10^{-2}$ we find that pebble accretion becomes only significant when the mass of the planet approaches $e^3 \tau_{\rm s} M_\star \approx 10^{-2} M_\oplus$.  But from the above discussion we know that gas drag and type I damping are ineffective to reduce the random motions of $400$-km-sized planetesimals.
In addition, a high planetesimal  inclination further reduces pebble accretion when it becomes larger than the aspect ratio of the pebble disk ($i_{\rm rms}> h_{\rm peb}$) \citep{Levison2015}. In \fg{eccinc} we find that this happens at $t\simeq 5 \times 10^{5}$ yr. This means after that the planetesimals exceed the pebble layer during part of its orbits, reducing the amount of pebbles they can ``eat''.

  \begin{figure}[t!]
    \sidecaption    
      \includegraphics[scale=0.49, angle=0]{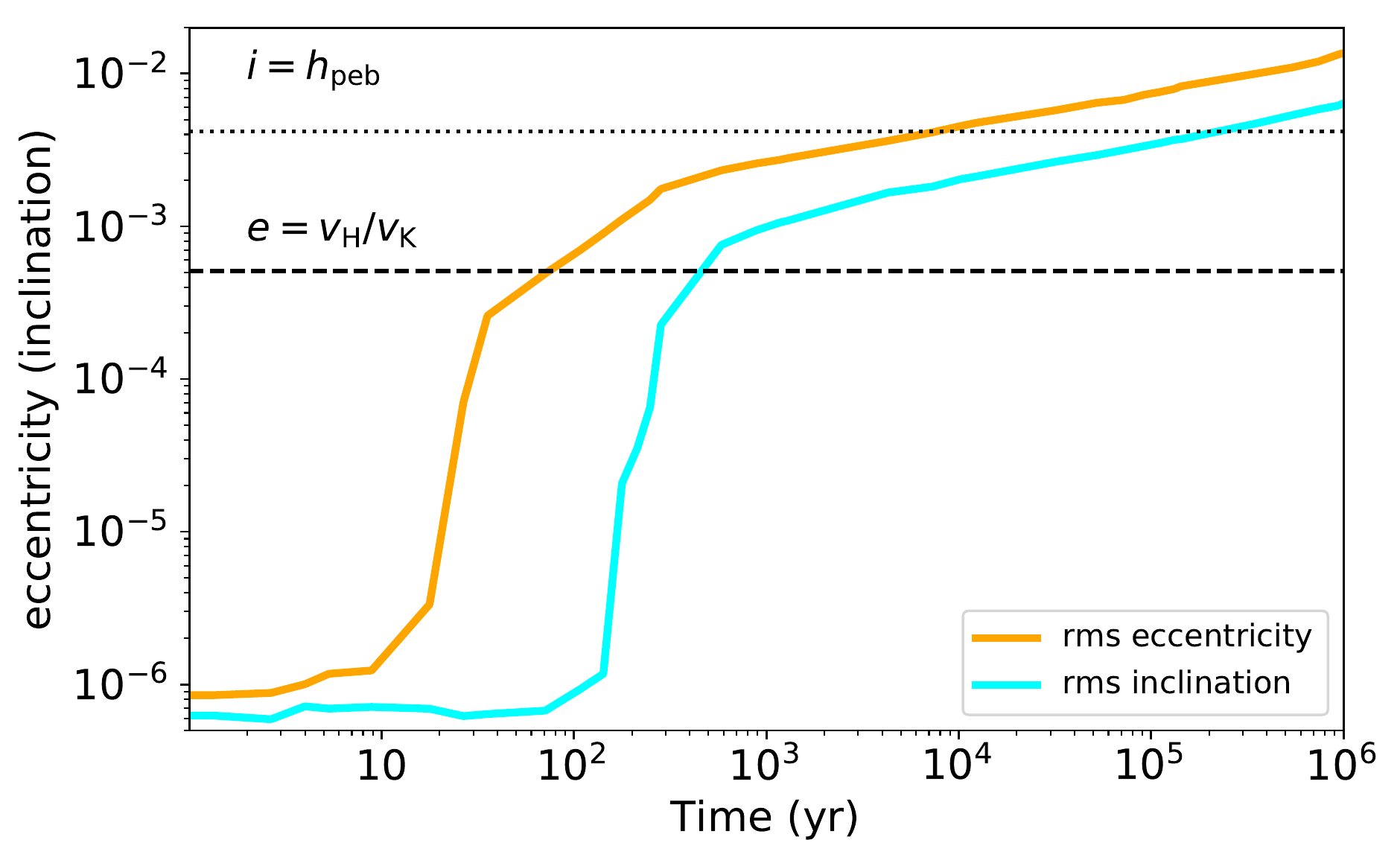}
      \caption{Root-mean-square (rms)  eccentricity (orange) and inclination (cyan) of planetesimals as functions of  time in \textit{run\_md\_1}.  The dashed line indicates the separation between the shear-dominated and dispersion-dominated regime, $e = v_{\rm H}/v_{\rm K}$. The dotted line indicates the inclination equal to the  aspect ratio of the pebble disk.  }
\label{fig:eccinc}
\end{figure} 
 \begin{figure}[t!]
    \sidecaption    
      \includegraphics[scale=0.49, angle=0]{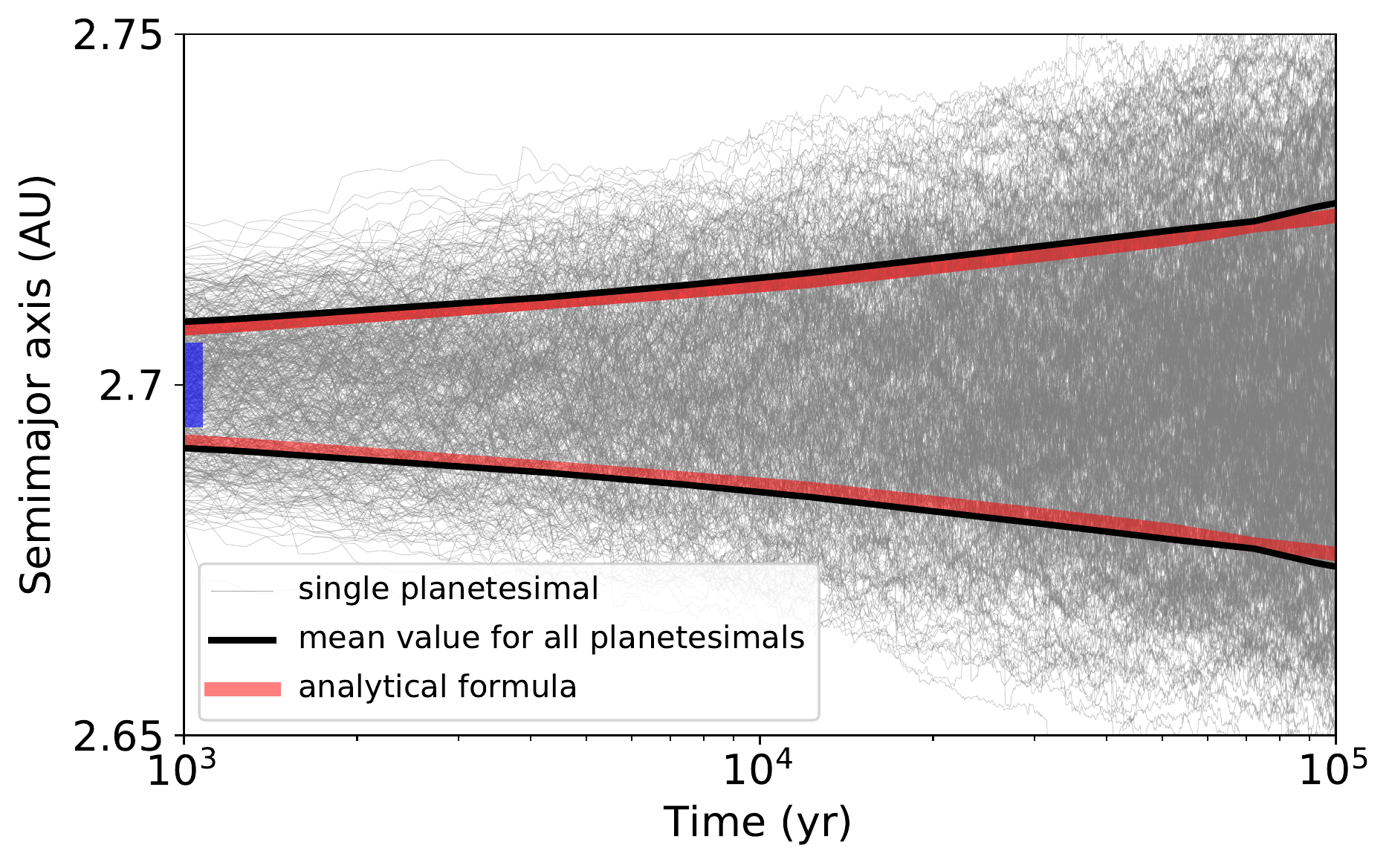}
      \caption{Orbital expansion  of planetesimals as a function of time. The grey and black lines are the semimajor axis of each planetesimal and the mean value of the planetesimal population, respectively. The red line denotes the analytical  expression for the width of the planetesimal belt from \Eq{width}.}
\label{fig:expansion}
\end{figure} 
 
\subsubsection{Spreading of the planetesimal belt}
\label{sec:spread}
In \fg{expansion} we plot the time evolution of the planetesimals' semimajor axes  in  $\it run\_md\_1$. We define the full width of the planetesimal belt as $ \Delta W = 2\sqrt{  \sum_{i} (a_{i} -r_{\rm ice})^2/N}$. The initial width is $ \eta  r_{\rm ice} $ (blue area). After their injection, planetesimals in this compact zone experience mutual gravitational interactions; scatterings excite their eccentricities and, in order to conserve the angular momentum, the range of the semi-major axes also expands. In \fg{expansion} we find that the width increases steadily with time and becomes $0.05$ AU after $10^{5}$ yr. 

This orbital expansion can be described by a diffusion process. \cite{Ohtsuki2003} obtain an analytical expression of the viscosity due to the mutual gravitational scattering based on equal size planetesimals. Substituting their viscosity ($\nu$ in their Eq.~(18)) into $\Delta W = \sqrt{\nu t}$,  we have   
 \begin{equation}
  \Delta W =  C_{\rm w} \sqrt  [\leftroot{-1}\uproot{2}\scriptstyle 3] {\left( N   \Omega_{\rm K} t /e^2  \right) } R_{\rm H}^2/a,
  \label{eq:width}
\end{equation} 
where  the prefactor $C_{\rm w}^3 = C_{\rm fit}  [24 { I(\beta)/\pi^2\ln(\Lambda ^2 + 1)]} $, $I(\beta)=0.2$,  $t$ is the time, $\Lambda=  \bar{i}_ {\rm rms}( \bar{e}_{\rm rms}^2 +  \bar{i}_{\rm rms}^2)/3$, and  $ \bar{i}_{\rm rms} = a i_{\rm rms}/R_{\rm H}$ and $\bar{e}_{\rm rms} = a e_{\rm rms}/R_{\rm H}$. The numerical factor of $C_{\rm fit} = 30$ is calibrated with our numerical simulations and is $\simeq 3$ times larger than \cite{Ohtsuki2003}'s result. This formula is obtained under the assumption that planetesimals are in the dispersion-dominated accretion regime. 
We show in \fg{expansion} that our analytical expression (\eq{width}) agrees well with the simulation (black line).
For fixed total mass $Nm_{\rm p}$ and $a$, $\Delta W $ increases with time ($\propto t^{1/3}$), the initial size of the planetesimal ($\propto R_0$), and decreases with the eccentricity ($e^{-2/3}$).

Since the ring belt expands over time, the surface density of the planetesimal also decreases. Therefore, the accretion of the planetesimals  formed from a narrow ring  will be longer than the classical /oligarchic accretion that assumes  an `infinite' width of the planetesimal disk \citep{Kokubo2000}.

 It is clearly seen from \fg{2d3d}b that the growth of planetesimals is slower than \fg{2d3d}a when considering the non-zero inclinations. We find that the runs in Table 1 with different initial randomness have a similar growth trend. The masses of the most massive planetesimals from \textit{run\_md\_1} to \textit{run\_md\_3} are all far below an Earth mass.

In conclusion, when starting from a narrow ring belt of $400$-km-sized planetesimals,  the growth of planetesimals is suppressed by the self-excitation of the planetesimals. Planetesimal accretion is slow and pebble accretion is inefficient. Formation of planeterary embryos, let alone super-Earth planets, is impossible in this scenario, even though $100 \Me$ of pebbles drift through. 

\subsection{Poly-dispersed initial conditions with large planetesimals}
\label{sec:poly-dispersed}

\begin{figure}[t]
    \sidecaption    
      \includegraphics[scale=0.49, angle=0]{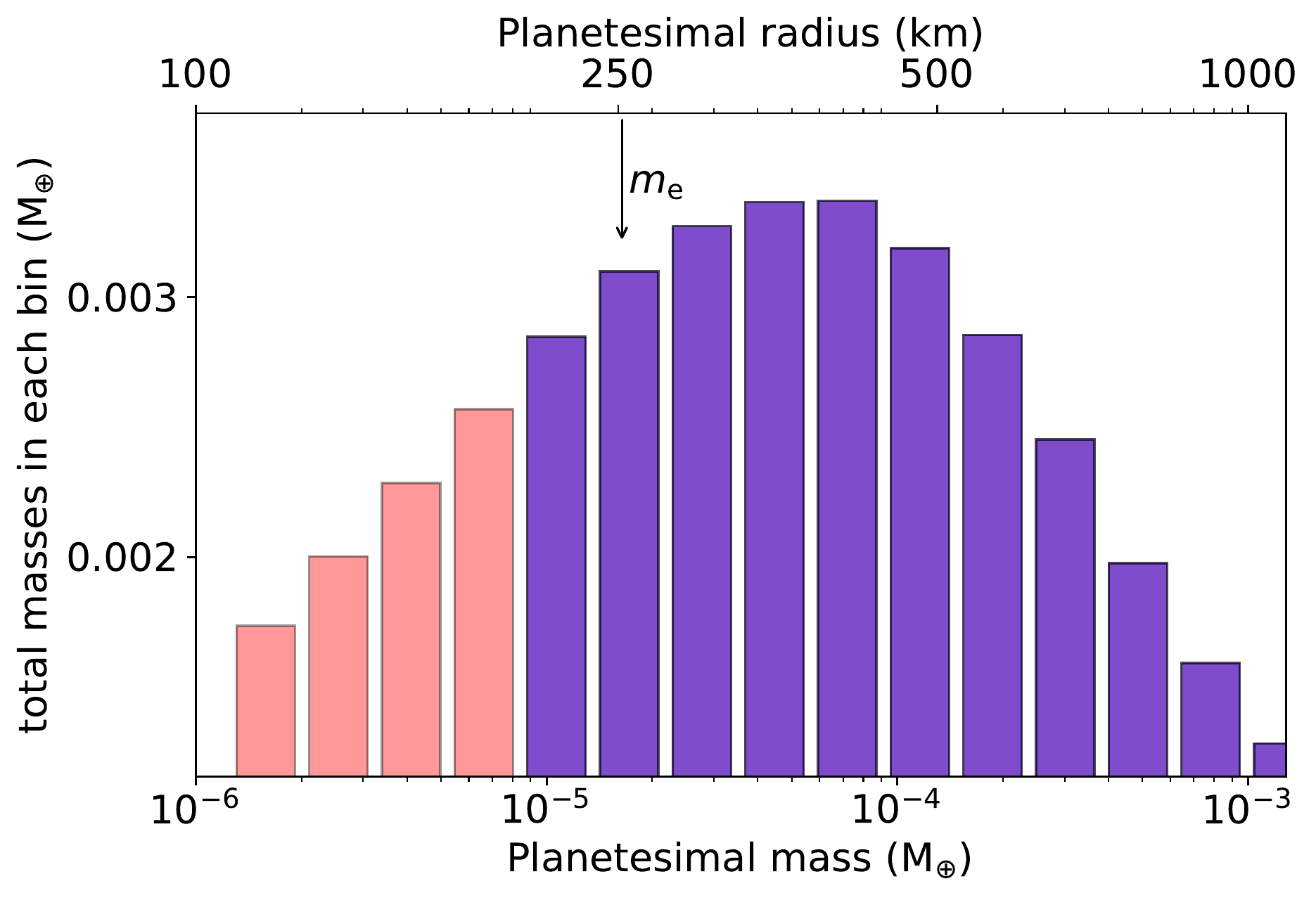}
      \caption{  Sum of planetesimal mass in each bin for all mass planetesimals (red + blue) and for the  simulated mass branch  (blue)  vs the initial planetesimal mass (size) from \eq{SIdist}. The initial mass function of planetesimals was based on streaming instability simulations with a minimum mass of $10^{-6} \Me$ ($100 $ km) and a characteristic mass of $1.6 \times 10^{-5} \Me$ ($ 250 $ km).
      We discrete the this distribution into different mass bins.  
  }
\label{fig:SImass}
\end{figure}

\begin{figure}[t]
    \sidecaption    
      \includegraphics[scale=0.7, angle=0]{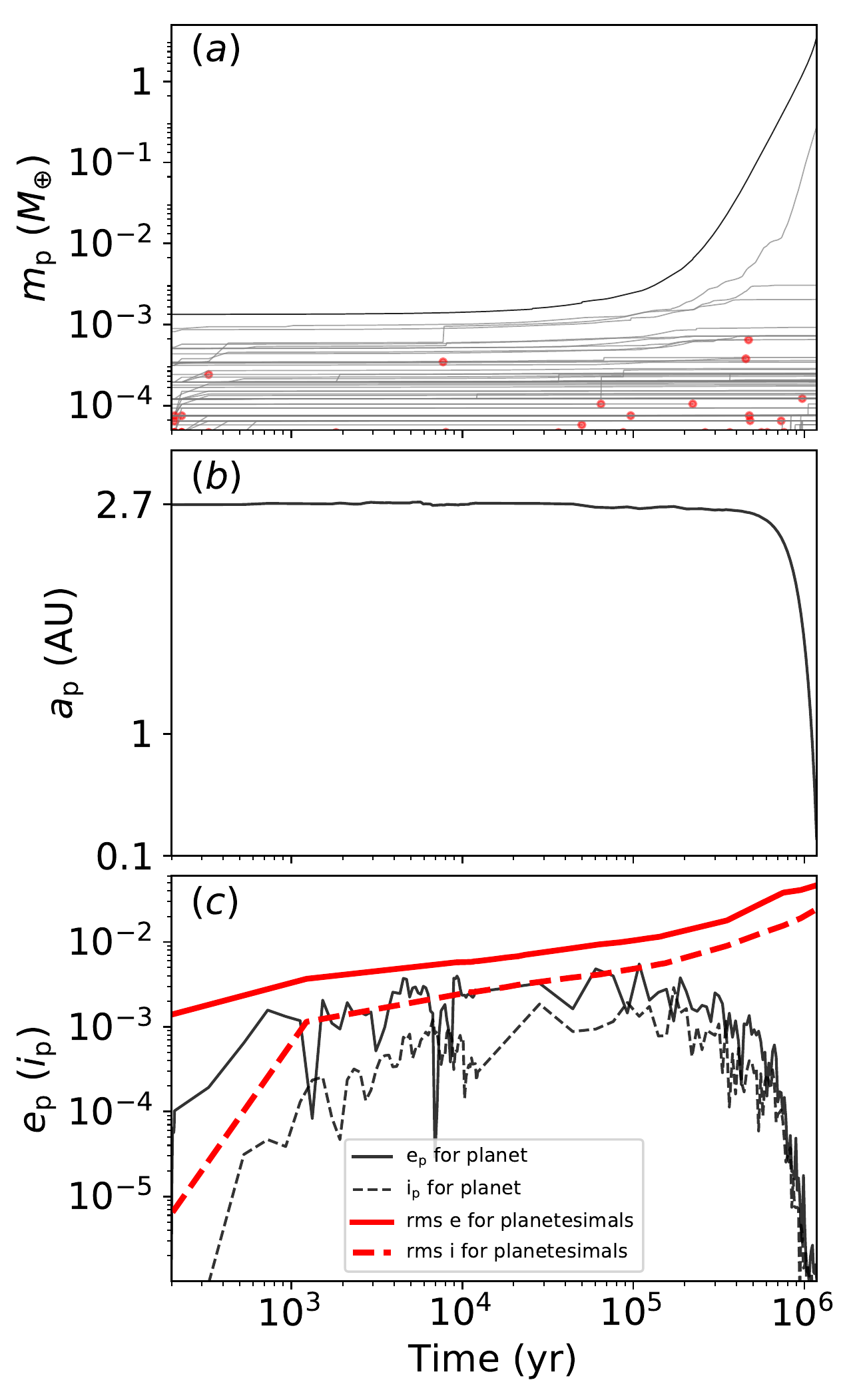}
      \caption{ Panel (a): mass growth of the planetesimals for a poly-dispersed size population of planetesimals based on the blue branch of \fg{SImass}. The black line represent the growth of the final most massive body and grey lines are for other planetesimals.  Red dots represents the merging of two bodies, marked at the position of the less massive body. The sudden growth is caused by planetesimal accretion and the smooth growth by pebble accretion. Note that $f_{\rm set}$ approaches $1$ when the planet is $0.01 \Me$. Panel (b): semi-major axis evolution of the final most massive body. Panel (c): Eccentricity (solid) and Inclination (dashed) evolution where black lines represent the final most massive body and the red lines indicate rms values of the planetesimals. The simulation is  from $run\_pd\_1$.}
\label{fig:poly-dispersed}
\end{figure}

In the previous section we found that when the sizes of planetesimals are all $400 \ \rm km$, their growth is inefficient at the ice line. Then the question is: what is the realistic sizes of the planetesimals?  In this section we consider a poly-dispersed population of planetesimals  from the streaming instability simulations.  
In this case, planetesimals evolve into two-component mass distribution (one large embryo + small planetesimals). The large body is able to start  rapid pebble accretion to eventually grow into a massive planet.  

The initial mass function of planetesimals by the streaming instability has been investigated by numerous authors in detail. \cite{Simon2016,Simon2017} find that the initial size can be described by a single power law.  Recently \cite{Schafer2017} suggests that this initial mass distribution is better fitted by a power law plus an shallow exponential decay. The  power law is for low and intermediate size planetesimals, while the exponential decay tail represents the high mass cutoff of planetesimals (Fig.~4 in \cite{Schafer2017}).
From  \cite{Schafer2017} simulations, the number fraction of planetesimals with $m_{\rm p}>m$ is given by their Eq.(19),
 \begin{equation}
  \frac{N_{>}(m)}{N_{\rm tot}} = \left( \frac{m}{ m_{\rm min}}  \right)^{-p} 
  \exp\left[  \left( \frac{m_{\rm min}}{m_{\rm e}} \right)^\beta  - \left( \frac{m}{ m_{\rm e} } \right)^\beta  \right].
  \label{eq:SIdist}
\end{equation} 
Following this distribution, we adopt their  parameters: $p = 0.6$, $\beta = 0.35$.  The minimum mass and  exponential characteristic mass are fitting parameters, which are given as  $m_{\rm min} =10^{-6} \Me$ ($100$ km), $m_{\rm e}$$=1.6\times10^{-5} \Me$ ($250$ km), respectively.  \Fg{SImass} illustrates the mass distribution of planetesimals for all mass range (red + blue) and for  the simulated mass branch (blue).  There are in total $0.039\Me $ planetesimals. Here we simulate the most massive $470$ bodies from this population, corresponding to the two-thirds of the total mass (blue brach in \fg{SImass}). The simulated bodies  span two order of magnitude in mass and a factor of $5$ in size,  from $200$ km to $1000$ km.  For simplicity, we neglect the lower one-thirds of small planetesimals. These bodies, if presented, would only modestly contribute to the planetesimal accretion and dynamical friction.

Other initial conditions are chosen to be the same as  the simulation of equal $400$-km-sized planetesimals (Table 1).  The mass growth of all planetesimals is shown in \fg{poly-dispersed}a while the semi-major axis of the largest body is shown in  \fg{poly-dispersed}b.  In \fg{poly-dispersed}c the eccentricity (solid) and inclination (dashed) evolution of the largest body is represented by the black line and the rms values of the planetesimals by the red line.

For the poly-dispersed population of  planetesimals simulated here,  most mass is in medium-sized planetesimals of $300$-$ 400$ km in size (\fg{SImass}).  Energy equipartition leads to a configuration that the massive planetesimals have low random velocities while the low-mass planetesimals have high random velocities  \citep{Ida1990,Goldreich2004}.  Therefore, the random velocity of the massive planetesimal is lower than the rms velocity dispersion due to dynamical friction.
At the beginning, the largest planetesimal (black line in \fg{poly-dispersed}a) can accrete more neighbouring planetesimals. Since its eccentricity and inclination remains low,  once the mass  is beyond  a  few $ 10^{-3} \Me$, the largest planetesimals will begin to accrete pebbles very efficiently.
 The mass of the largest planetesimal then increases further, rendering pebble accretion more efficient compared to lower-mass planetesimals. Therefore, this massive embryo dominates the subsequent growth and dynamical evolution of the system.

 Rather than stop the simulation at $1$ Myr, we continue it until the most massive planet migrated into the inner $0.1$ AU ( the assumed location of the inner disk edge) in this illustrated run (\textit{run\_pd\_1}). 
 We find that the planet reaches the inner disk edge at $1.2$ Myr when the mass is $3.9 \Me$.  The second largest body only grow its mass rapidly when the largest one migrated out of the ring. In the end of the simulation the second largest body is still one order of magnitude less massive than the largest one.
 
To conclude, the presence of   $\sim 10^{3}$ km planetesimals in a sea of predominantly a few hundreds km bodies will significantly speed up planet formation. 
This poly-dispersed population results in a low velocity dispersion of the massive body.  Both effects (high $m_{\rm p}$ and low $\Delta v$) {facilitate the transition to the pebble accretion dominated phase (phase B)}  and promote the formation of a  super-Earth planet after $\simeq1$ Myr.

\subsection{Two-component initial conditions as a result of runaway growth of small planetesimals}
\label{sec:two-component}
As demonstrated above,  a poly-dispersed population of planetesimals containing initially a few very large bodies can form planets rapidly. Such an initial planetesimal mass distribution would be a consequence of forming planetesimals directly from the streaming instability.  Generally,  a lower pebble surface density yields smaller planetesimals in streaming instability simulations \citep{Johansen2015,Simon2016}. For the single-site planetesimal formation scenario, the pebble density is locally enhanced at a particular disk location (\eg , the ice line). But how strong the enhancement  is determined by complicated  physical processes such as water vapor diffusion and condensation \citep{Schoonenberg2017}. Taken these factors and uncertainties into account, we also consider the case of an initially small planetesimal size.  Nevertheless, we argue in \se{runaway} that in this circumstance runaway growth of planetesimals proceeds fast, producing the desired two-component mass distribution that promotes the formation of massive planets. 

\subsubsection{Runaway growth in a narrow planetesimal belt}
\label{sec:runaway}

The characteristic runaway planetesimal accretion timescale is \citep{Ormel2010b}
\begin{equation}
\begin{split}
  \tau_{\rm rg}  & =  C_{\rm rg} \frac{ \rho_{\bullet} R_0}{ \Sigma_{\rm p} \Omegak} 
  = 10^{5} \left(  \frac{C_{\rm rg}}{0.1}  \right)  \left(  \frac{R_0}{100 \rm \ km} \right) \\
  & \left(   \frac{\Sigma_{\rm p}}{10 \rm \ g\,cm^{-2}} \right)^{-1}  \left( \frac{a}{ 2.7 \AU } \right)^{3/2 }      \ \rm yr. 
  \label{eq:runaway}
 \end{split}
\end{equation} 
where $C_{\rm rg} $$ \simeq $$0.1$ is a numerical-corrected prefactor from \cite{Ormel2010b}, and $\rho_{\bullet} =1.5 \ \rm g\, cm^{-3}$ is the internal density of the planetesimal. 
Small planetesimals have three advantages in terms of  mass growth.
First, based on \eq{runaway} the runaway growth timescale is  proportional to the initial size of the planetesimals  ($R_0$). It takes less time to form a runaway body (before the onset of the oligarchic growth) when starting with smaller planetesimals.
Second, the eccentricity and inclination excitation by self-stirring is less severe for smaller planetesimals (\Eq{tau_vs}), which also boosts the planetesimal and pebble accretion. 
Third, as shown by in \se{spread}, the spreading of their semimajor axes  increases with the size of the planetesimal ($R_0$ $\propto$ $R_{\rm H}$ in \eq{width}). For the same total planetesimal mass, smaller planetesimals expands their orbits less, resulting in a higher surface density. Therefore, the subsequent accretion also becomes faster.    

Accounting for these effects, we assume that the planetesimals are all $100  \ \rm km$  in size ($\simeq 10^{-6} \Me$ in mass).  
These planetesimals would undergo a faster runaway growth and less orbital expansion compared to $400$-km- sized bodies. Since the runaway growth will end up with a steep mass distribution ($d N/dm_{\rm p}  \propto m_{\rm p}^{-2.5}$  \citep{Wetherill1993,Kokubo1996,Kokubo2000,Ormel2010b,Morishima2008}, most of the mass remains in small planetesimals. As discussed in \se{poly-dispersed}, the inclination and eccentricity of the large body could therefore be  damped by small planetesimals through dynamical friction. 
 The planet can accrete more pebbles when the orbit remains nearly coplanar and circular. 

The key difference between the classical runaway and oligarchic growth scenario and our scenario is that we here consider planetesimal formation at a single location. In this study we  propose a clean and simplistic physical picture.   The hypothesis is that \textit{after} the planetesimal runaway growth phase, the system can be well-described by a two components: a single big embryo of radius $\sim 10^{3}$ km  and a swarm of $100$ km size planetesimals, which dominate the total mass. We reasonably neglect bodies in between these two are dynamically insignificant and cannot compete with the runaway body in terms of growth (also shown in \fg{poly-dispersed}a).
Altogether, the situation resembles the classical runaway/oligarchy transition, except that, in our case, there is only one single ``oligarch''.

Why only one big embryo instead of multiple oligarchs? The transition from the runaway to the oligarchic growth takes place when the viscous stirring timescale $\tau_{\rm vs} $  (\eq{tau_vs}) equals  the runaway timescale $\tau_{\rm rg}$ (\eq{runaway}) \footnote{This criterion is different from \cite{Kokubo1998}. See discussion in \cite{Ormel2010b}. }. When $\tau_{\rm vs} > \tau_{\rm rg}$, the stirring of the random velocities is smaller compared to the accretion, which is the key feature of runaway growth. When $\tau_{\rm vs} < \tau_{\rm rg}$, the eccentricity growth is faster than the accretion, and  the growth transitions to the oligarchic regime. 
The transition size of the planet between two regimes is given by Eq. (11) of \cite{Ormel2010b},
 \begin{equation}
 \begin{split}
  R_{\rm rg/oli}  \simeq &  850 \left(\frac{C_{\rm rg}}{ 0.1}\right)^{3/7}  \left(\frac{R_0}{ 100 \ \rm km}\right)^{3/7} 
   \left(\frac{a}{ 2.7  \AU }\right)^{5/7} 
     \left(\frac{\Sigma_{\rm p}}{  1 \ \rm g \,cm^{-2}} \right)^{2/7}    \rm km. 
  \label{eq:Rtr}
   \end{split}
\end{equation} 
Substituting \eq{runaway} into \eq{width}, we obtain that when the massive body grows into a transition radius, the planetesimal belt width is   
\begin{equation}
  \Delta W  =  \sqrt{  \frac{C_{\rm rg} C_{\rm w}^3 R_{\rm H}^3}{2R_0}  }.
  \label{eq:width2}
\end{equation} 
 We note that $R_{\rm H}$ is the Hill radius of the planetesimal with size of $R_0$. 
From  $\Sigma_{\rm p} =  M_{\rm tot,pl}/2\pi  a \Delta W $  and  and  \eq{width2},  we calculate that the belt width extends to $\Delta W =0.029$ AU and the planetesimal surface density reduces to $\Sigma_{\rm p} = 2 \rm \ g\,cm^{-2}$ after the runaway growth ($\sim 10^{5}$ yr). 
Substituting the above valuses into \eq{Rtr}, we obtain, 
\begin{equation}
 \begin{split}
  R_{\rm rg/oli}  \simeq   1000 \left(\frac{C_{\rm rg}}{ 0.1}\right)^{3/7}  \left(\frac{R_0}{ 100 \ \rm km}\right)^{1/7}   \left(\frac{M_{\rm tot,pl}}{ 0.039 \Me}\right)^{2/7}  \  \rm km.
  \label{eq:Rtr2}
   \end{split}
\end{equation} 
We therefore conclude that the feeding zone of this embryo ($ \simeq 10 \ R_{\rm H}$, \cite{Kokubo1998}) is similar to the width of the planetesimal belt, supporting one single embryo assumption.

\subsubsection{The superparticle approach}
\label{sec:approach}
Since the total planetesimal mass is $0.039 \Me$  and small planetesimals dominate the mass, there are  in total  $35000$ 100-km-sized planetesimals ($\simeq$$10^{-6} \Me$) in this case. It is prohibitively computationally  to simulate the interactions among all these bodies with an N-body scheme.
Therefore, we adopt the superparticle approach to mimic the dynamics of small bodies (see \cite{Levison2015,Raymond2016} and references therein), in which $N_{\rm sp}$ small planetesimals are clustered as one superparticle. The superparticle still feels the aerodynamical gas drag  as if it was a single $100$-km-sized planetesimal, but the mass of the particle is $N_{\rm sp}$ times higher ($m_{\rm sp}= N_{\rm sp} m_0$). These superparticles gravitationally interact with the embryo but not with each other. This is  based on  the fact that the collision timescale among small planetesimals is much longer than the embryo-planetesimal collision.  In principle, the approach requires $m_0 \ll  m_{\rm sp} \ll m_{\rm p}$, where $m_{\rm p}$ is the embryo's mass. The latter inequality is needed to ensure that dynamical friction operates correctly.  The simulation would be very time-consuming for a too small $N_{\rm sp}$,  while  the dynamical evolution  may become artificially stochastic for a too large $N_{\rm sp}$ (low number of superparticles). 

We here assume that the velocity dispersion of small planetesimals is excited by the planetesimals  themselves, and that it is close to their escape velocity ($\delta v \simeq v_{\rm esc} = \sqrt{2G m_0/R_0}$),  \ie , $ e \simeq e_{\rm esc}= v_{\rm esc}/v_{\rm K}$. The eccentricities and inclinations are adopted to follow the Rayleigh distributions, where  $e_0 =2 i_0= e_{\rm esc}$. Their semi-major axes are randomly initialized  from $r_{\rm ice}(1-\eta/2)$ to $r_{\rm ice}(1+ \eta/2)$.       In addition, since pebble accretion is extremely inefficient for $100$ km size planetesimals (the inequality (\ref{eq:PA-condition}) is far from satisfied), it is  reasonable to neglect pebble accretion of the superparticles.  The embryo is initially placed at $r_{\rm ice}$.  

\subsubsection{Results}
\label{sec:sp_result}

We have conducted simulations with different masses of the superparticle, $N_{\rm sp}=25$ and $50$, respectively. The results are in good agreement with each other (see Appendix for details). The results presented in \ses{sp_result}{sp_timescale} are based on the $N_{\rm sp}=50$ run.

\begin{figure}[tbh!]   
   \includegraphics[scale=0.6, angle=0]{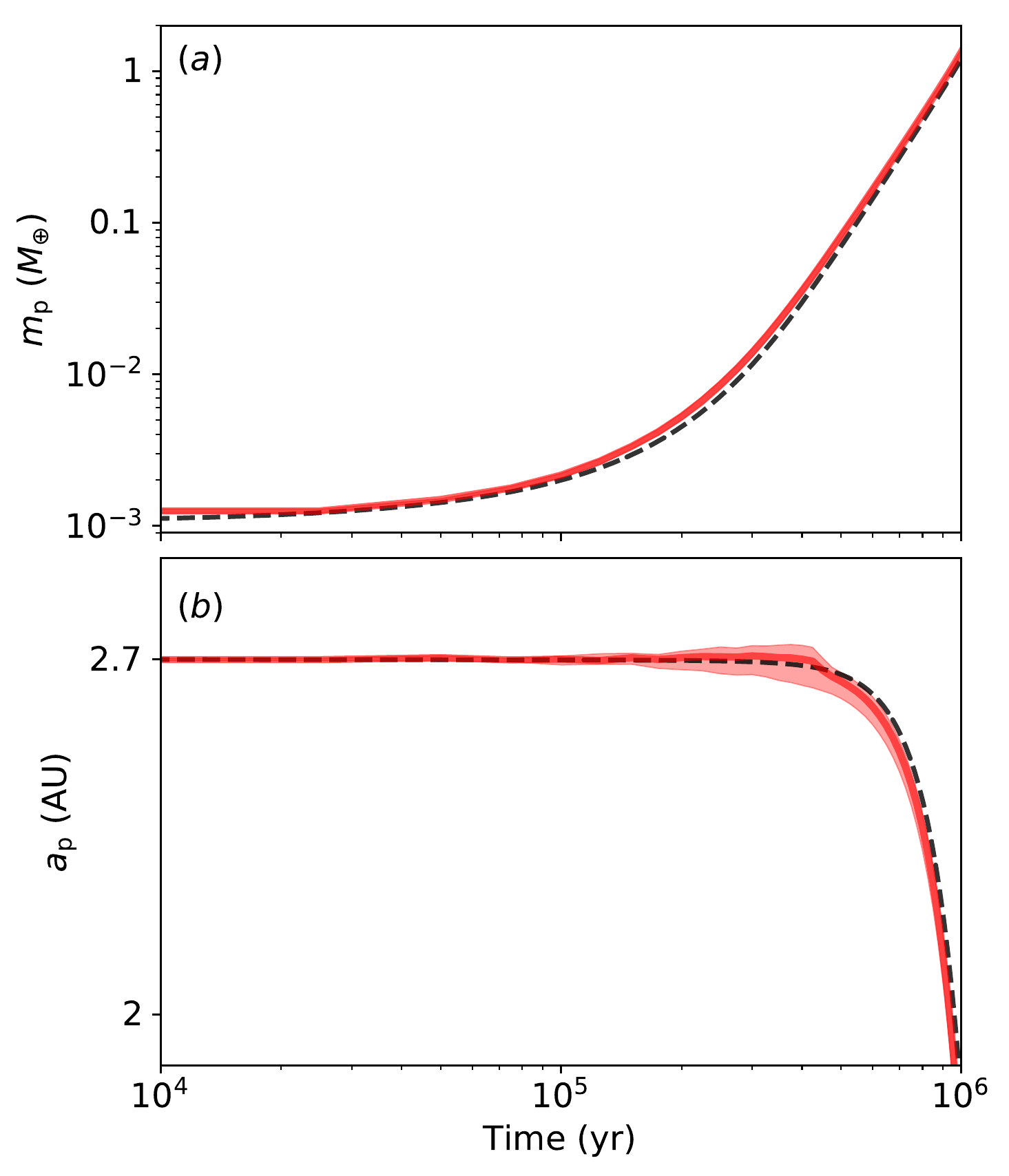}
    \caption{
        Panel (a): Mass growth  of the embryo for the two-component initial conditions in $run\_sp$.  Panel (b): Semi-major axis evolution of the embryo. The red thick line is for the averaged value and the light red is for the spreading among  three different runs.  The black dashed line is for the growth of a single embryo on a circular and coplanar orbit without any planetesimal accretion.  } 
\label{fig:runaway}
\end{figure} 

The disk and pebble parameters in this (\textit{run\_{sp}}) are taken to be identical to \textit{ run\_{pd}} in \se{poly-dispersed}. The mass of the embryo is chosen to be the transition value  ($ m_{\rm p} = 10^{-3} \Me$ when $R_{\rm p} = R_{\rm rg/oli}$  in  \eq{Rtr2}).
Three individual simulations (\textit{run\_{sp}\_1} to \textit{run\_{sp}\_3}) are conducted, varying only the initial positions and orbital phase angles of the planetesimals. In addition, we also conduct a separate simulation for the growth of a single embryo without small planetesimals. This comparison enables us to isolate the effect of N-body dynamics (planetesimal accretion). Simulations are terminated when $t = 1$ Myr.

\Fg{runaway} shows the mass growth and orbital evolution of the embryos in panel (a) and (b), respectively. 
For the superparticle approach, the thick red lines represent the mean values averaging three runs (\eg , $ {\bar m} (t) =\sum_{i=1}^{3} m_{\rm i}(t)/3$), whereas the light red areas represent the spreading among these runs.   It can be seen  that the difference in mass and semi-major axis among the three simulations is  small.   The black dashed line represents the growth of a single embryo (assuming it is on a circular and coplanar orbit) purely by pebble accretion. We find that when starting with the transition size embryo, the planetesimal accretion are already modest and the mass growth is mainly driven by pebble accretion.

When the mass of the embryo is beyond $0.1 \Me$, the type I migration becomes important.  The planet would take  $\sim 0.5$ Myr to migrate into the inner region of the disk. 
Eventually, a $1.3 \Me$  planet forms within $ 1$ Myr.

\subsubsection{Timescale analysis}
\label{sec:sp_timescale}
In this section we analyse the growth timescale of the embryo. 
The pebble accretion timescale in the 2D regime is given by
\begin{equation}
\begin{split}
 \tau_{\rm PA,2D} = & \frac{m_{\rm p}}{  \varepsilon_{\rm PA,2D}  \dot M_{\rm peb}} \simeq 9 \times 10^{4} f_\mathrm{set}^{-1}  \left( \frac{ m_{\rm p}}{0.05 \Me} \right)^{1/3}  \\
 & \left( \frac{ \taus}{0.1} \right)^{1/3}
  \left( \frac{\eta}{2.5 \times 10^{-3}}  \right)  \left( \frac{\dot M_{\rm peb}}{  100 \rm \ M_{\oplus}/Myr}  \right)^{-1} 
  \rm \ yr.
  \label{eq:tau_PA2d}
\end{split}  
\end{equation}
where we use \eq{eps-2D} and $\Delta v$ is assumed to be dominated by the Keplerian shear since the eccentricity of the embryo is  insignificant due to dynamical friction. From \eq{fset0} the above $f_\mathrm{set}$ evaluates  as
\begin{equation}
 \begin{split}
    f_\mathrm{set} 
    &= \exp{\left[-0.07   \left(\frac{\eta}{2.5\times 10^{-3}} \right)^2   \left(\frac{m_{\rm p}}{0.01 \Me} \right)^{-2/3} \left( \frac{\taus}{0.1} \right) ^{2/3}   \right]}.
\label{eq:fset}
  \end{split}
\end{equation}
When the planet is $\gtrsim 10^{-2} \Me$, $f_{\rm set} \simeq 1$.  
From \eq{fset}, we clearly see that  $f_{\rm set}$ becomes smaller when a planet is less massive  or their pebble-planet relative velocity is higher. 

Similarly, in the 3D limit (\eq{eps-3D}), the pebble accretion timescale becomes
 \begin{equation}
 \begin{split}
 \tau_{\rm PA,3D} = & \frac{m_{\rm p}}{  \varepsilon_{\rm PA,3D}  \dot M_{\rm peb}} \simeq 9 \times 10^{4} f_\mathrm{set}^{-2}  \left( \frac{ h_{\rm peb}}{ 4.2 \times 10^{-3}} \right)  \\
 &   \left( \frac{\eta}{2.5 \times 10^{-3}}  \right)  \left( \frac{\dot M_{\rm peb}}{  100 \rm \ M_{\oplus}/Myr}  \right)^{-1} 
  \rm \ yr.
  \label{eq:tau_PA3d}
  \end{split}  
\end{equation}
From the planet mass dependence on  \eqs{tau_PA2d}{tau_PA3d}, we know that the pebble accretion is in $3$D  when the planet  mass is low, and it transitions to $2$D accretion when the planet becomes more massive. The transition mass between $2$D and $3$D is $\simeq 0.05 \Me$ for the adopted parameters.

In order to numerically capture the transition from the planetesimal dominated accretion to the pebble dominate accretion (phase A to phase B), we conduct the superparticle approach simulations starting with a less massive embryo ($m_{\rm p} =3\times 10^{-4} \Me$) here.
The other conditions are  the same as $ run\_{sp}$  in Table 1.  Again, three individual superparticle approach simulations are performed with randomly varied initial positions and orbital phase angles of the planetesimals.

These timescale are shown in \fg{analysis}, where the grey,  blue,  light blue and  magenta  lines represent growth timescales obtained from the simulation,  the pebble accretion in $3$D (with and without $f_{\rm set}$) and in $2$D, respectively. We can identify the following different growth stages.

When the mass of the embryo is lower than $\sim 10^{-3} \Me $, the mass growth is dominated by planetesimal accretion (phase A).  In this phase the growth time is relatively long, and decreases with the growing mass of the big body. 
Since the timescale calculation is affected by the stochasticity of the collisions, the average value among three different runs (grey line) is given in \fg{analysis}.  The pebble accretion is inefficient  in this phase because of the low embryo's mass (small $ m_{\rm p}$ in \eq{tau_PA3d}) and hence the weak setting accretion (small $f_{\rm set}^2$  in \eq{tau_PA3d}). When the mass of the embryo becomes larger, $3$D pebble accretion is more efficient than planetesimal accretion (phase B). The transition mass from the planetesimal accretion dominated  regime to the pebble accretion dominated regime is  close to $10^{-3} \Me$. With increasing mass $f_{\rm set}$ increases. When $f_\mathrm{set}$ approaches unity, the growth timescale becomes independent of the planet mass (see \eq{tau_PA3d} and the light blue in \fg{analysis}). Finally,  growth enters  the $2$D  pebble accretion regime when the mass approaches $0.1 \Me$. In that case growth timescale starts to increase,  $\tau_{\rm PA,2D} \propto m_{\rm p}^{1/3}$ (magenta line in \fg{analysis}), since $\varepsilon_{\rm PA,2D}$ no longer scales linearly with the planet mass.

\begin{figure}[t!]
    \sidecaption    
         \includegraphics[scale=0.5, angle=0]{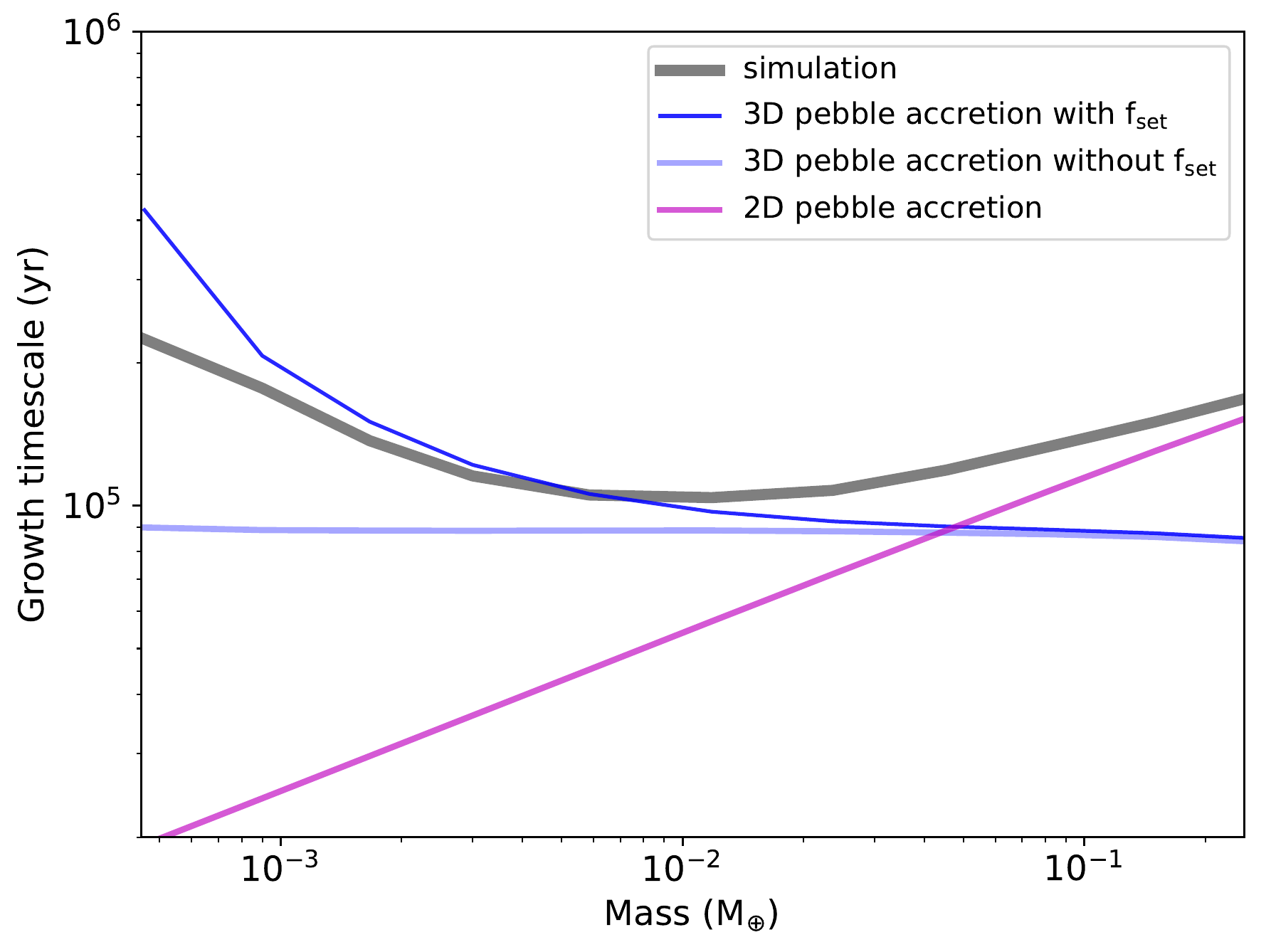}
      \caption{ 
          Growth timescale as a function of  the embryo's mass.  The grey line represents the simulation whereas  the blue and magenta lines are timescales for the analytical $3$D and $2$D pebble accretion. The  dark (light) blue line corresponds to $\tau_\mathrm{3D}$ including (or excluding) the   $f_{\rm set}$ factor  (\eq{tau_PA3d}). } 
\label{fig:analysis}
\end{figure}

\subsection{Discussion} 
 
Comparing the cases of initial $100$-km-sized and $400$-km-sized planetesimals, our result suggests that starting with smaller planetesimals is more optimal for forming planets.  The reason is as follows: 1) the runaway body's size is insensitive to the initial planetesimal size ($R_{\rm rg/oli} \propto R_0^{1/7}$, \eq{Rtr2}), 
and 2) since the orbital expansion is less significant for smaller planetesimals ($\Delta W \propto R_0 $ from \Eq{width2}) and $\Sigma_{\rm p} \propto \Delta W^{-1}$,
 the runaway accretion timescale is strongly correlated with the initial size ( $t_{\rm rg} \propto R_0^2$ from \eq{runaway}). 
For the mono-dispersed population of $400$-km-sized planetesimals, planet growth fails mainly because  it takes too long to evolve into a two-component mass distribution. However, in the case of $100$-km-sized planetesimals, this configuration can be realized within a fraction of the disk lifetime. In that case, the eccentricity and inclination of the large embryo remains low due to the dynamical friction of  small planetesimals. It is therefore possible for the largest embryo to accrete pebbles efficiently and grow into an Earth-mass planet. As a result, massive planets could form from planetesimals with a smaller initial size. 

The above discussed preferable size of the planetesimal ($R_0$) to grow protoplanets is  dependent on the formation location of planetesimals. When the planetesimals form further out, from \eq{runaway} the runaway planetesimal growth timescale increases. On the other hand, the starting size of the embryo for the efficient pebble accretion also increases with the distance  \citep{Visser2016}. This is because $f_{\rm set}$ becomes smaller when  $\eta$ increases with $r$.  Therefore, both two effects further limit the growth of the protoplanets from a large size planetesimal (\eg, $400$ km) when the planetesimal formation occurs at a large distance.

Our findings are similar to \cite{Levison2015}. They report that large core formation by pebble accretion is only feasible after a period of classical planetesimal runaway growth. The system then ends up with a few large embryos and a swarm of small planetesimals. The embryos stir the random velocities of all small planetesimals, preventing any further growth by pebble accretion. 
In their case, planetesimals were born over the entire disk region and  assumed to  follow the same power-law as the disk gas,  which is based on the classical planet formation scenario (\eg , \cite{Ida2004a} and \cite{Mordasini2009}).  The key difference with this work is that we consider the planetesimals to form at a single site disk location with a narrow radial width. Nevertheless,  the `common ground' between both works is that a two-component mass distribution for planetesimals is essential to promote further growth from planetesimals into protoplanets.

\begin{table*}
    \centering
    \caption{Simulations for the parameter study in \se{parameter}. The last two columns represent the final mass and the time when the planet reaches the inner $0.1$ AU.  }
    \begin{tabular}{lllllllllll|}
        \hline
        \hline
        Name  (description)  &  $\dot M_{\rm peb}$ ($\rm M_{\oplus}\,Myr^{-1}$) & $\alpha_{\rm t}$  &  $\taus$   &  $\dot a/a$ & $m_{\rm p,f} (\Me)$ & $ t_{\rm f} (\rm \ Myr) $  \\
      \hline
   \textit{run\_{fid}} (fiducial) & $100$ & $10^{-3}$ & $0.1$  &  yes & $4.0$ & $1.2$   \\
    \textit{run\_nmig} (no migration) & $100$ & $10^{-3}$ & $0.1$   &  no &$3.8$ & $1.2$    \\
    \textit{run\_alpha} (low turbulence) & $100$ & $10^{-4}$ & $0.1$  &  yes  &$4.1$ & $0.9$  \\
    \textit{run\_tau} (low Stokes number) & $100$ & $10^{-3}$ & $0.03$   &  yes  & $4.8$ & $1.5$ \\
    \textit{run\_hpeb} (high pebble flux) & $200$ & $10^{-3}$ & $0.1$  &  yes  & $6.8$ & $0.7$  \\
    \textit{run\_lpeb} (low pebble flux) & $50$ & $10^{-3}$ & $0.1$   &  yes   & $2.3$ & $2.1$ \\
         \hline
        \hline
    \end{tabular}
    \label{tab:tab1}
\end{table*}

\section{Growth and migration in the pebble accretion dominated regime}  
\label{sec:parameter}  
In this section, we particularly focus on growth in the pebble accretion dominated regime (phase B).
We start the mass of the embryo from $10^{-3} \Me$.  As illustrated in \fg{runaway},  the pebble accretion dominates the growth and the planetesimals' contribution can be neglected after $m_{\rm p} \gtrsim 10^{-3} \Me$.  The relative velocity of the massive embryo is very low  due to the dynamical friction by small planetesimals and later on  type I tidal damping when it grows larger. It is therefore justified to assume the embryo is on a circular and coplanar orbit.
We conduct simulations for a single embryo and neglect all planetesimals, investigating the role of various disk and pebble parameters as shown in Table 2.  Simulations are terminated when the planet has migrated inside of 0.1 AU.

In Table 2,  the fiducial run  (\textit{run\_{fid}}) is adopted to be the same  disk and pebble values  as in the previous sections. From \textit{run\_{nmig}} to \textit{run\_{lpeb}}, only one parameter is varied  compared to the fiducial \textit{run\_{fid}}.  
For instance, in \textit{run\_nmig} we assume the planet is at the zero-torque location, where $\dot a/a =0$. In this case, the planet does not undergo type I migration. Therefore, the planet accretes materials \textit{in-situ} at the ice line. For a comparison between \textit{run\_nmig} and \textit{run\_fid}, we will gain a knowledge of the effect of migration on the growth of the planet. For this purpose, we stop \textit{run\_nmig} at the same time when the planet  in \textit{run\_fid} migrates inside of $0.1$ AU.
For a comparison between \textit{run\_fid}  and the other individual runs (\textit{run\_alpha}, \textit{run\_tau},  \textit{run\_hpeb},  \textit{run\_lpeb}) , we can understand the effect of disk turbulence, pebble size and pebble mass flux on the planet growth.

We find  in \fg{parameter}  that in a low turbulent disk (\textit{run\_alpha}, magenta) and in a disk with high pebble flux (\textit{run\_hpeb}, thick red), the mass growth is faster than the fiducial disk (red). In these two circumstances the planets reach $4.1 \Me$ and  $6.8 \Me$, respectively, when they migrate inside of $0.1$ AU within $1$ Myr.  We find the mass growth in \textit{run\_nmig} is very similar to the fiducial run but slightly less efficient.   At $t = 1.2$ Myr, the final planet mass is  $4.0 \Me$  in \textit{run\_fid} while  in \textit{run\_nmig} the planet always stays at the ice line and attains $3.8 \Me$. 
It is also clearly seen that when the pebble flux is lower (\textit{run\_lpeb}, light red), 
or the Stokes number is  lower (\textit{run\_tau}, orange), the mass growth is slower than the fiducial run. The protoplanet with $m_{\rm p} \gtrsim 1 \Me$ does not form within $1$ Myr from these two configurations.  In  \textit{run\_tau} the planet attains $4.8 \Me$ when it arrives at the inner edge of the disk at $t = 1.5$ Myr, while in  \textit{run\_lpeb} it grows into a  $2.3 \Me$ planet at $t = 2.1$ Myr.

The planet grows much faster in a high pebble flux disk.  The effect of pebble mass flux on planet growth is intuitive since pebble accretion benefits from a high pebble flux  (a massive pebble disk).   When the planet reaches $6.8 \Me$  and migrates into the inner disk cavity at $t = 0.7$ Myr in \textit{run\_hpeb}, the planet mass is only $0.02 \Me$  in \textit{run\_lpeb}.  We find  a super linear correlation between the planet mass ($m_{\rm p}$)  and  the integrated pebble flux ($\dot{M}_\mathrm{peb} t$).
 A factor of $4$ change in  $\dot M_{\rm peb}t$ results in more than two order of magnitude growth in $M_{\rm p}$. 
 This is due to the fact that pebble accretion efficiency ($\varepsilon_{\rm PA}$) increases with $m_{\rm p}$   in both $2$D and $3$D regime. The planet can initially accrete more pebbles in a high pebble flux disk. It becomes more massive due to this faster growth and would further accrete even a higher fraction of pebbles.  This positive feedback promotes the rapid growth of the planet in a high pebble flux disk. 

When the disk turbulence  is lower, the pebble scale height becomes smaller (\eq{Hpeb}).  The transition mass  from $3$D  to $2$D pebble accretion  will also become lower. Therefore, in \textit{run\_alpha} the embryo starts efficient $2$D pebble accretion early than the fiducial run. Eventually, it takes less time to grow into a super-Earth planet in a lower turbulent disk.

The Stokes number has an  effect opposite to the turbulent $\alpha_{\rm t}$ in terms of the pebble scale height.  Lower Stokes number pebbles mean  they are more tightly coupled to the gas, and the pebble scale height becomes larger.  Therefore,  from \eqs{tau_PA2d}{tau_PA3d} transition mass  from $3$D to $2$D in \textit{run\_tau} is $0.15 \Me$,  three times higher than the fiducial run. In this case starting with $10^{-3} \Me$, the planet grows a significant fraction of its mass in the slow, $3$D pebble accretion regime. On the other hand, when the planet enters $2$D pebble accretion, the accretion is faster when the Stokes number is lower (\eq{tau_PA2d}). Balancing these two effects, we find that the planet growth is slightly slower  in \textit{run\_tau}  compared to the fiducial run.

 The effect of pebble mass flux on the planet growth is super linear. A high pebble disk mass benefits the formation of a massive planet. A less turbulent disk and a large Stokes number pebbles also promotes pebble accretion and  the formation of a massive planet.

\begin{figure}[tbh!] 
    \includegraphics[scale=0.5, angle=0]{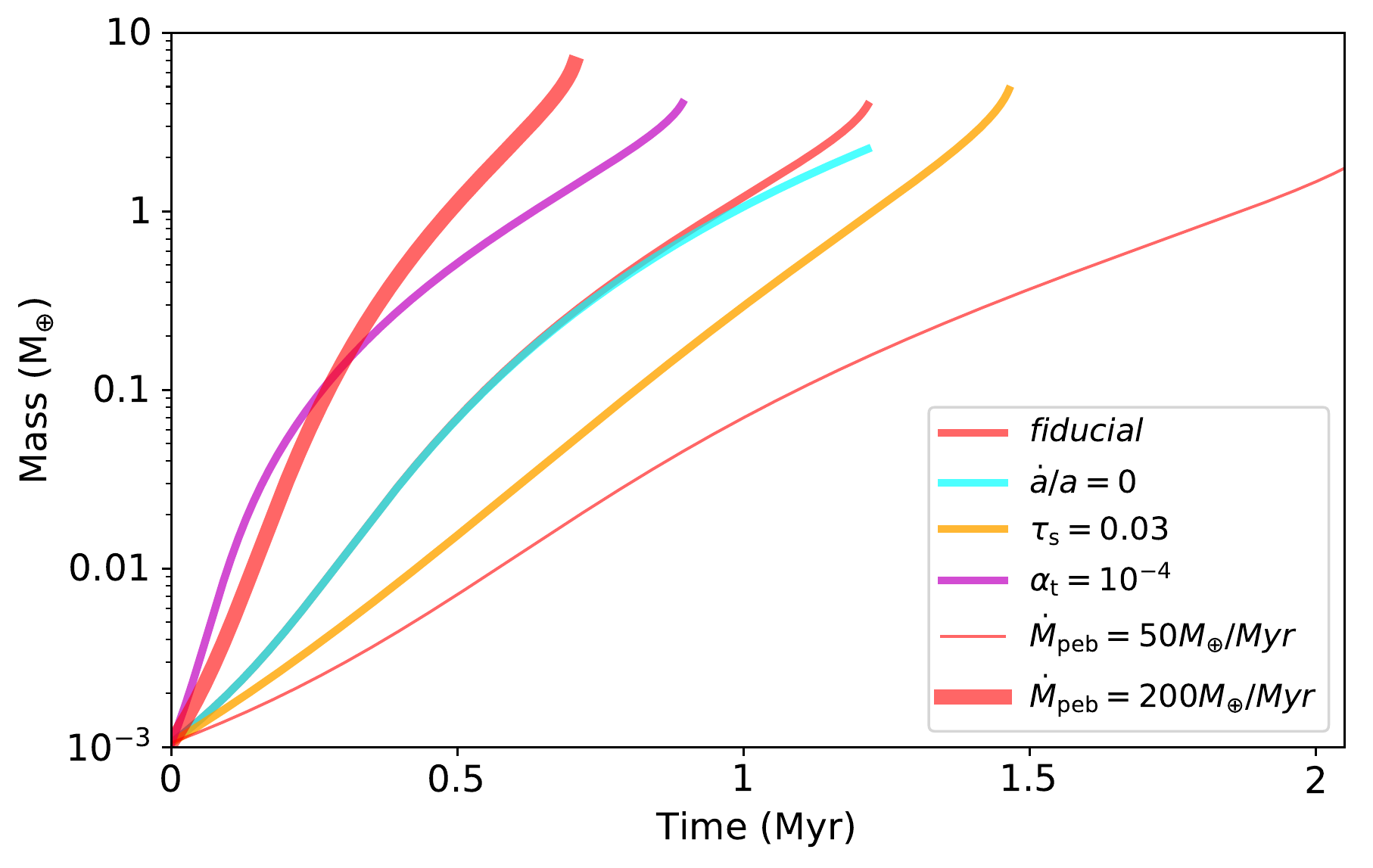}
    \caption{
        Mass growth of one single embryo  in \se{two-component} for different runs given in Table 2.  The red, cyan, orange and magenta corresponds to \textit{run\_fid},  \textit{run\_nmig}, \textit{run\_tau} and \textit{run\_alpha}, respectively. The thick and thin red lines represent  \textit{run\_hpeb} and \textit{run\_lpeb}.       
       }
\label{fig:parameter}
\end{figure}

\section{Summary}
\label{sec:summary}
Streaming instability is an important mechanism to convert pebbles into planetesimals. It occurs at the location in the disk where the pebble density is locally enhanced  (\eg , the ice line). 
In this work, we have focused on the growth of the planetesimals after they have formed by this mechanism at a single site disk location.

This annular (ring) planetesimal formation scenario differs from the classical planetesimal accretion scenario from two aspects.  First, streaming instability generates planetesimals in a narrow ring  at a specific location. Most of the mass is in the largest planetesimals of a few hundred km in size. In contrast, in the classical scenario  planetesimals form everywhere in the disk, \eg, typically following the surface density distribution of the gas.  In simulations their planetesimal surface density remains constant by suppling material from the neighbouring region \citep{Kokubo2000}. This means that the orbital spreading is unimportant, in contrast to the ring formation scenario.  Second, the planetesimals  in our scenario grow their mass by accreting surrounding planetesimals {\it and} inwardly drifting pebbles. The mass at which the planet transitions from accreting predominantly planetesimals to pebbles occurs at  $\simeq 10^{-3}\Me$. 

We have modified the \texttt{Mercury} N-body code to perform  simulations of the mass growth and orbital evolution of these planetesimals (\se{method}).  The code includes the effects of gravitational interactions and collisions among planetsplanetesimals, planet-disk interactions (gas drag and type I torque), pebble accretion based on the calculation of \cite{Liu2018} and \cite{Ormel2018} that accounts for the disk  parameters ($\taus$, $\eta$, $\alpha$ and $\dot M_{\rm peb}$) and planet properties ($m_{\rm p}, a, e, i$). Simulations with different initial planetesimal sizes  and disk parameters are investigated.

The key findings of this study are the following:
\begin{enumerate}
    \item  Protoplanets cannot emerge  from a mono-dispersed population of $400$ km size planetesimals, fuelled by a $100 \Me$ reservoir of pebbles in the outer disk. Although the initial eccentricities and inclinations are very tiny,  they   soon get excited through gravitational scatterings. Mechanisms such as gas drag, type I damping are not sufficient enough to to damp their random velocities.
   Both planetesimal and pebble accretion are strongly suppressed when inclinations and eccentricities of planetesimals become moderate.  In this circumstance, the growth of the planetesimals is mainly in the slow planetesimal accretion dominated phase (\se{mono-dispersed}).   
    \item  Protoplanets can form when streaming instability has \textit{in addition} spawned a population of larger planetesimals. The largest body grows by planetesimal accretion. Soon after it approaches $ 10^{-2} \Me$, the growth enters the rapid pebble accretion dominated regime. During this time the random velocity of the largest body remains low through the dynamical friction of small planetesimals.  Finally a super-Earth planet can form within $1$ Myr (\se{poly-dispersed}).   
    \item Alternatively, protoplanets also form out of their birth ring when the initial size of the planetesimals is small (\eg , $100$ km).  These small planetesimals  are expected to undergo a runaway planetesimal accretion to form a massive embryo rapidly. In this way, the two-component mass distribution is also achieved. We find that an Earth mass planet can form as well (\se{two-component}) 
    \item  Planets grow larger when the pebble mass flux is higher, the disk is less turbulent, the Stokes number of pebbles are larger.  In particular, the growth of the planet mass increases super linearly with the disk pebble mass flux (\se{parameter}).       \end{enumerate}

Planetesimal accretion and pebble accretion are not two isolated processes in planet formation. From the streaming instability point of view, pebbles are converted into planetesimals, whereafter these planetesimals accrete nearby planetesimals and  pebbles at the same time. The total amount of solids in disks is either in small pebbles, or in large planets/planetesimals. Therefore, a certain level of competition exist between planetesimal formation and planet growth by pebble accretion.   

In addition, this work only considered the case of a single burst of planetesimal formation by streaming instability. In reality, even at the ice line the streaming instability may be triggered multiple times (episodic bursts) during the gas disk lifetime, because the planets migrate out of their birth ring. Protoplanets then emerge sequentially and a chain of multiple planets forms, as envisioned by \cite{Ormel2017b}.  

To address these issues, a global disk model is needed. Using a novel Lagrangian approach, such a model has just been developed \citep{Schoonenberg2018a}. Because of the flexibility of the Lagrangian (particle-oriented) model, it is straightforward to couple it to the N-body model presented in this paper. Such a model can then, potentially, simulate planet formation in its entirety -- starting from dust coagulation and ending with a planetary architecture. It can be applied to model ``complete'' (as far as one can tell) planetary systems, such as those discovered around TRAPPIST-1 \citep{Gillon2017}.

\appendix
\section{Convergence test for superparticle simulations}
\label{sec:appendix}

We show the simulations with different mass of the superparticle ($m_{\rm sp} = N_{\rm sp} m_{\rm 0}$) for $run\_sp$ starting with $m_{\rm p} = 3 \times 10^{-4} \Me$ in \se{sp_timescale}. Three simulations with randomlized initial conditions are performed for each set of $N_{\rm sp}$.  The results are shown in \fg{compare} where red, green represent  $N_{\rm sp} = 25$ and $50$, respectively. The thick line represents the mean value averaged from three individual simulations, for instance, $ {\bar m} (t) =\sum_{i=1}^{3} m_{\rm i}(t)/3$ whereas the light region marks the range between the minimum and maximum values from the three simulations.
 
\fg{compare} shows the mass, semi-major axis evolution of the embryo and  eccentricities of both planetesimals and the embryo. We find  that the mass and semi-major axis converge quite well for the above two $N_{\rm sp}$.   In \fg{compare}a the difference between the mean mass  for the above $N_{\rm sp}$s is smaller compared to the spreading among individual runs (stochastic N-body effects). The excitation of planetesimals by the presence of the embryo is also agreed with each other in in \fg{compare}c.

 To summarize, results from the above three tested $N_{\rm sp}$ are in general agreement with each other. 

\begin{figure}[tbh!]    
   \includegraphics[scale=0.7, angle=0]{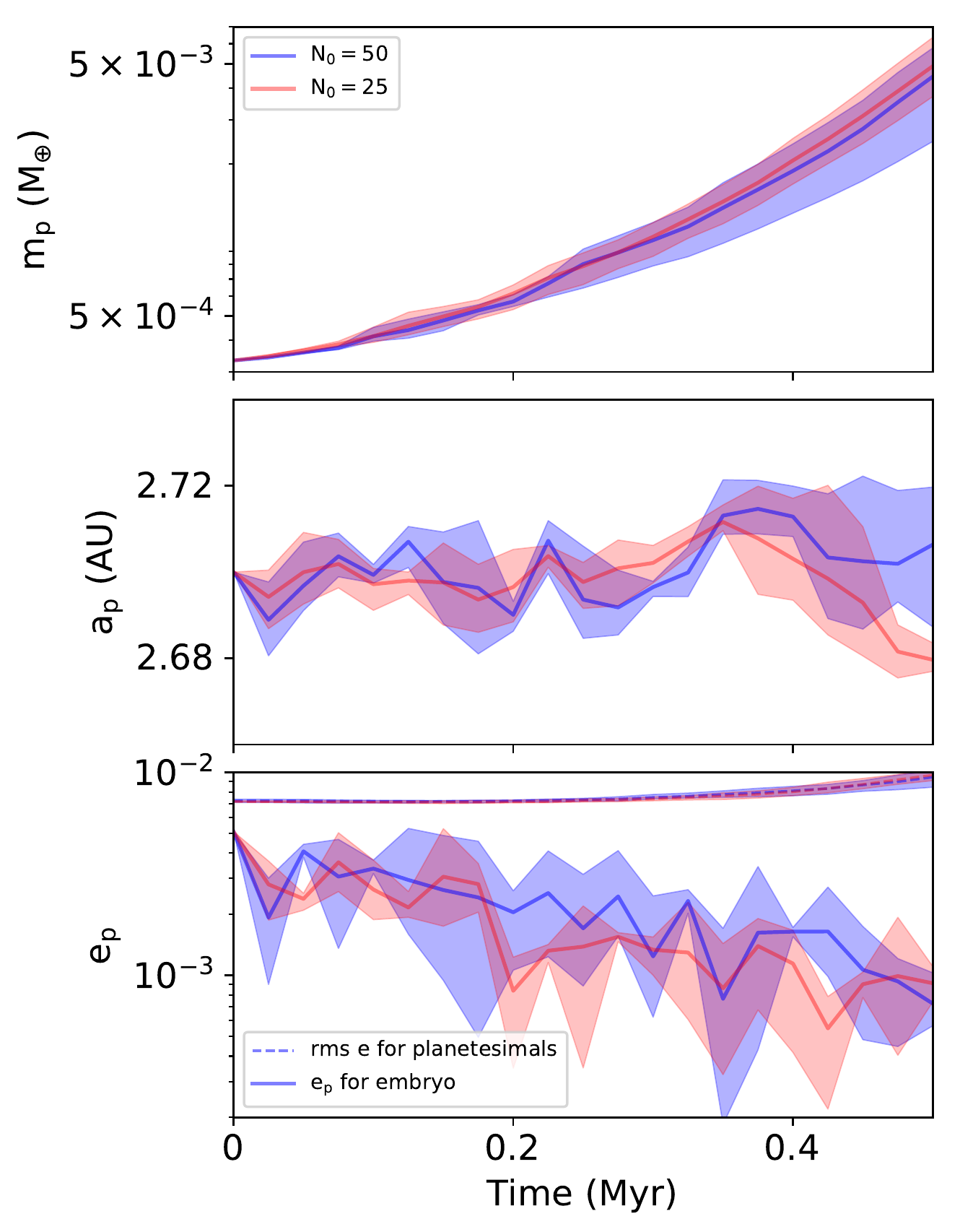}
    \caption{
  Convergence test for $N_{\rm sp} =25$ (red) and $50$ (blue).  Mass, semi-major axis and eccentricity evolution of the embryo are shown. The line represents the mean value of three individual runs while the area indicates the scattering from these runs.  In the bottom panel the RMS eccentricity of the planetesimals are shown in the dashed line.  
      }
\label{fig:compare}
\end{figure}

\begin{acknowledgements}
We thank Chao-Chin Yang, Lixin Li, Carsten Dominik, Melvyn Davies for useful discussions, and Bertram Bitsch, Djoeke Schoonenberg for proofreading the manuscript. We also thank the anonymous referee for their insightful suggestions and comments. 
Beibei Liu thanks the support of  the Netherlands Organization for Scientific Research (NWO; VIDI project 639.042.422), the European Research Council (ERC Consolidator Grant 724687-PLANETESYS) and the Swedish Walter Gyllenberg Foundation.
  Chris Ormel is supported by the Netherlands Organization for Scientific Research (NWO; VIDI project 639.042.422).  
Anders Johansen is funded by the Swedish Research Council (grant 2014-5775), the Knut and Alice Wallenberg Foundation (grants 2012.0150, 2014.0017) and the European Research Council (ERC Consolidator Grant 724687-PLANETESYS). 
The computations are performed on resources provided by the Swedish Infrastructure for Computing (SNIC) at the
LUNARC-Centre in Lund.

\end{acknowledgements}

\bibliographystyle{aa}
\bibliography{reference}

\end{document}